\documentclass[12pt,english,reqno]{smfart}

\usepackage[table]{xcolor}

\usepackage[utf8]{inputenc}
\usepackage{subfig}
\usepackage{mathrsfs}
\usepackage{amsthm}
\usepackage{amscd}
\usepackage{amstext}
\usepackage{amsfonts}
\usepackage{amssymb}
\usepackage{graphicx}
\usepackage{setspace}
\usepackage{esint}
\usepackage{rotating}
\usepackage[nocompress]{cite}

\usepackage[all]{xy}
\usepackage{array}

\usepackage{hyperref}
\hypersetup{colorlinks=true,linkcolor=magenta,anchorcolor=green,citecolor=cyan,filecolor=black,menucolor=black,urlcolor=brown}
\makeatletter

 \usepackage[compat=1.1.0]{tikz-feynman}
\usepackage{tikz,contour}
\usetikzlibrary{calc,decorations.markings}
\pgfdeclarelayer{bg}    % declare background layer
\pgfsetlayers{bg,main}  % set the order of the layers (main is the standard layer)

\def\su{\circleddash}
\numberwithin{equation}{section}
\numberwithin{figure}{section}
\theoremstyle{plain}
\newtheorem{thm}{\protect\theoremname}[subsection]
  \theoremstyle{plain}
  \theoremstyle{definition}
  
  \theoremstyle{plain}
  
  \theoremstyle{remark}

  \theoremstyle{plain}
 
 \newtheorem{conj}[thm]{\protect\conjecturename}
  \providecommand{\corollaryname}{Corollary}
  \providecommand{\definitionname}{Definition}
  
  \providecommand{\remarkname}{Remark}
\providecommand{\theoremname}{Theorem}
\providecommand{\lemmaname}{Lemma}
\providecommand{\conjecturename}{Conjecture}
\title[Algorithms for minimal Picard-Fuchs operators]{
Algorithms for minimal Picard-Fuchs operators\\ of Feynman integrals
}

 \author[P. Lairez]{Pierre Lairez}
 \address{Inria, Uni. Paris-Saclay, Palaiseau, France}
\author[P. Vanhove]{Pierre Vanhove}
 \address{
Institut de Physique Th\'eorique, Universit\'e Paris-Saclay, CEA, CNRS, F-91191 Gif-sur-
Yvette Cedex, France}

\thanks{IPHT-T/t22/058}
\date{\today}

\begin{document}

 \begin{abstract}
%   {\bf Draft version \today\ at \thistime}
   In even space-time dimensions the multi-loop Feynman integrals are
   integrals of rational function in projective space. By  using
   an algorithm that extends the Griffiths--Dwork reduction for
   the case of projective hypersurfaces with singularities, we derive
   Fuchsian linear differential
   equations, the Picard--Fuchs equations, with respect to kinematic
   parameters for a large class of massive multi-loop Feynman
   integrals. With this approach we obtain the differential operator for
   Feynman integrals to high multiplicities and high loop orders. Using recent factorisation algorithms we give the minimal
   order differential operator in most of the cases studied in this paper.  Amongst our results are that the order
   of Picard--Fuchs operator for the
   generic massive two-point $n-1$-loop sunset integral in two-dimensions is
   $2^{n}-\binom{n+1}{\left\lfloor \frac{n+1}{2}\right\rfloor }$
   supporting the conjecture that the  sunset Feynman integrals are
   relative periods  of Calabi--Yau of dimensions $n-2$. We
   have checked this explicitly till six loops. As well,
   we obtain a particular Picard--Fuchs operator of order 11 for the
   massive five-point tardigrade non-planar two-loop integral
 in four dimensions  for
generic mass and kinematic configurations, suggesting that it arises from $K3$
   surface with Picard number 11.  We determine as well
Picard--Fuchs operators of
two-loop graphs with various multiplicities in four dimensions,
finding Fuchsian
differential operators with either Liouvillian or elliptic solutions.
\end{abstract}
\maketitle
\newpage
\tableofcontents
\newpage
%%%%%%%%%%%%%%%%%%%%%%%%%%%%%%%%%%%%%%%%%%%%%%%%%%%%%%%%%%%%%%%%%%
\section{Introduction}

Feynman integrals enter the evaluation of many physical observable quantities in
particle physics, gravitational physics, statistical physics, and
solid-state physics. They are multi-valued functions, with non-trivial
monodromies. They present branch
cuts (associated with particle production) and their analytic or numerical
evaluations are challenging.
The identification of the kind of special functions needed to evaluate
the Feynman integrals is difficult question under study from the early
days of Quantum field theory~\cite{Golubeva,Pham} and is still an
active field of research e.g.~\cite{Panzer:2015ida,Duhr:2019wtr,Mizera:2019ose}.

Broadhurst and Kreimer~\cite{Broadhurst:1995km,Broadhurst:1996kc}
remarked that the definition of the Feynman integrals resembles the definition of period integrals given by
Kontsevich and Zagier in~\cite{Kontsevich:2001}. Bloch, Esnault, and
Kreimer~\cite{Bloch:2005bh} and Brown~\cite{BrownCosmic}  showed that the
Feynman integral is a relative period integral of the mixed Hodge structure
determined by the graph polynomials defining the Feynman
integrals.   There are increasing evidence that some Feynman integrals are
relative periods integrals of (singular) Calabi--Yau geometries~\cite{Brown:2009ta,Bloch:2014qca,Bloch:2016izu,Bourjaily:2018ycu,Bourjaily:2019hmc,Bourjaily:2018yfy,Klemm:2019dbm,Bonisch:2020qmm,Bonisch:2021yfw,Bourjaily:2022bwx,Forum:2022lpz,Duhr:2022pch}.

One aim of this work is to sharpen this correspondence by
deriving Picard--Fuchs differential operators associated to the Feynman
integrals by studying the case where a Feynman integral $I_\Gamma(t)=
\int_{\Delta_n}  \Omega_\Gamma(t)$ is
given by the integral over the positive orthant $\Delta_n$, defined
in~\eqref{e:DefDomain}, of  rational differential form in projective space
\begin{equation}\label{e:GenericRat}
  \Omega_\Gamma(t)= {\mathcal U(\underline x)^{n-(L+1)D/2}\over (
   \mathcal F_0(\underline x) +t \mathcal F_1(\underline x))^{n-LD/2}} \Omega_0^{(n)}
\end{equation}
where $n$, $L$ are positive integers and $D$, the space-time
dimension, is taken to be an even positive integer $D$, such that
$n-LD/2>0$ for convergence of the integral. We have set 
$\underline x:=[x_1:\dotsb:x_n]\in\mathbb P^{n-1}$, and $\Omega_0^{(n)}$ is
the canonical differential form on $\mathbb P^{n-1}$, given in~\eqref{e:Omega0def}.
The homogeneous polynomial $\mathcal U(\underline x)$
has degree $L$ and the homogeneous polynomials $\mathcal
F_0(\underline x)$ and $\mathcal F_1(\underline x)$ have degree
$L+1$ so that the integrand is a well-defined rational differential
form.  We will restrict to the case of  converging integrals.

Our approach uses an implementation of the Griffiths--Dwork reduction
algorithm adapted to the case of period of rational
integrals with non-isolated singularities. We 
construct a Fuchsian differential operator in the variable~$t$ annihilating
the integrand
\begin{equation}\label{e:LQ}
 \left(\mathcal L_t-\mathcal C(t,\partial_t;\partial_{\underline
  x},\underline x)\right)  {\mathcal U(\underline x)^{n-(L+1)D/2}\over (
   \mathcal F_0(\underline x) +t \mathcal F_1(\underline x))^{n-LD/2}} =0\,.
\end{equation}
This annihilator is composed of
 a  Picard--Fuchs differential  operator  acting only on the parameter $t$ 
\begin{equation}
  \mathcal L_t= \sum_{r=0}^{o_n} p_r(t) 
  \left(d\over d t\right)^r
\end{equation}
and a certificate part
\begin{equation}\label{e:Cdef}
  \mathcal C(t, \partial_t;\partial_{\underline
  x},\underline x)=\sum_{i=1}^n  {\partial\over
  \partial x_i} Q_i(t,\partial_t;\partial_{\underline x},\underline x)
\end{equation}
with
  \begin{equation}\label{e:Q}
        Q_i(t,\partial_{t};\partial_{\underline x},\underline x)=\sum_{0\leq
    a\leq o_i'}\sum_{0\leq b_j\leq \tilde
    o_j\atop 1\leq j\neq n} q_{a,b_1,\dots,b_r}^{(i)}(t;\underline x)
  \left(\partial\over\partial t\right)^{a}\prod_{j=1}^n
  \left(\partial\over\partial x_j\right)^{b_j}.
  \end{equation}
which  are  differential operators acting on the parameter $t$ and
  the integration variables $\underline x$.   From the
  identity~\eqref{e:LQ} we deduce that the Feynman
  integral~\eqref{e:GenericRat} satisfies the  inhomogeneous
  differential equation
  \begin{equation}\label{e:PFgeneric2}
    \mathcal L_t I_\Gamma(t)= \mathcal S(t) .   
  \end{equation}
The inhomogeneous term $\mathcal S(t)  $ is a sum of the $n$ contributions evaluated on the
boundary components arising from the 
evaluation  of the action of certificate on the integrand on the
boundary of positive orthant. Each boundary components are rational
differential form for Feynman integral in $\mathbb P^{n-2}$ (see~\cite{Vanhove:2014wqa}).  For fixed
values of degree of homogeneity $L$ and $D/2$, we are naturally led to
study the rational periods integrals~\eqref{e:GenericRat}, with
increasing values of $n$.  The inhomogeneous term in the
right-hand-side of~\eqref{e:PFgeneric2} arises by
integrating the certificate in~\eqref{e:Cdef}, and for its derivation
involves the blow-up needed to define the Feynman integral as a
relative period integral. Since this
work is primarily focused on the derivation of the homogeneous
differential operator in the left-hand-side of~\eqref{e:PFgeneric2} we
will not need to specify the blow-up procedure needed for a proper
definition of the variation of mixed Hodge structure associated with a
given Feynman integral~\cite{Bloch:2005bh,BrownCosmic}.

This work has two main aims. (1) The first one is to provide a convenient
tool for exploring the relation between Feynman integrals and periods
integrals. From the knowledge of the Picard--Fuchs operators~$\mathcal L_t$  we get an information (an upper bound) on the number of
independent periods associated with the differential form~\eqref{e:GenericRat}. We check if the Picard--Fuchs
operator is factorisable and when  possible
using the factorisation algorithm implemented in~\cite{facto}. The
irreducibility of the Picard--Fuchs operator implies that the
operator is minimal, but there are factorisable operators that are
minimal as we will see in the case of the kite case in Section~\ref{sec:kite}.
The analytic form of the Picard--Fuchs operators and their
singularities provide an important hint about the algebraic geometry
for which the Feynman integral is a relative period integral.  We will
find Picard--Fuchs operators suggestive of rational surfaces, elliptic
curves, $K3$ surfaces,
Calabi--Yau $n$-folds. By working with
increasing values of $n$ we  determine if new types of periods arise
from the homogeneous differential operator.
(2)  The second aim is to provide an efficient tools for 
deriving the differential equation for Feynman integrals numerically. The  present
algorithms give a way to analysis relatively high loop (we will study
cases up to $L=6$ loops order) and high
multiplicities (we will study cases with  up to seven external states)  integrals.

The application of these techniques to   Feynman integrals is
described in~\cite{VanhoveISSAC}. But the implementation of the
Griffiths--Dwork reduction algorithm is complicated by the presence of
non-isolated singularities.
For the case of the two-loop sunset integral, one can apply the standard
Griffiths--Dwork reduction algorithm for deriving the Picard--Fuchs
operator (see~2.3.2 of~\cite{Vanhove:2018mto}).
But unlike the two-loop sunset integral, Feynman
integrals have generically non-isolated singularities, preventing a direct use of
the Griffiths--Dwork reduction.
We will use an extended version of the Griffiths--Dwork
formalism adapted to the case 
with non-isolated singularities
that implement the pole reduction in the integrand using
Syzygies~\cite{Lairez}.\footnote{A Magma implementation is available at~\href{https://github.com/lairez/periods}{github.com/lairez/periods}.}

In this work, we will first recall the relation between the Feynman
integrals and relative period integrals in Section~\ref{sec:Feynman}.  
In Section~\ref{sec:PFdef} we describe the algorithm we will be using
for deriving the Picard--Fuchs operators. In Section~\ref{sec:sunset}
we derive the Picard--Fuchs operators for the sunset family of integral
up to six-loop order with generic mass configurations. We show the
factorisation of the Picard--Fuchs operator 
when specialising the mass configurations.  In
Section~\ref{sec:icecream} we consider the multi-scoop ice-cream cone
graphs. Adding an external momentum to the multi-loop sunset graph yields a multi-scoop ice-cream cone graph. We show how this affects the
differential operator. In Section~\ref{sec:twoloop} we give the differential operator
for various two-loop integrals. We show the triviality of the kite
integral in two dimensions. In four dimensions, we will obtain that the massive kite
and massive
double-box integral differential operator is order 2 compatible
with an elliptic curve, and that the massive tardigrade differential
operator is suggestive of a $K3$ surface with Picard number 11 for
generic mass and kinematic configurations.
We conclude with Section~\ref{sec:conclusion}.

We have collected various numerical results and expressions that are
too long for being reported in this text on the following online pages
\begin{itemize}
\item The~\href{https://nbviewer.org/github/pierrevanhove/PicardFuchs/blob/main/Identity.ipynb}{identity} in eq.~\eqref{e:Fid} 
\item The differential operators for the sunset graph at~\href{https://nbviewer.org/github/pierrevanhove/PicardFuchs/blob/main/PF-2sunset.ipynb}{2 loops},
\href{https://nbviewer.org/github/pierrevanhove/PicardFuchs/blob/main/PF-3sunset.ipynb}{3 loops}, 
\href{https://nbviewer.org/github/pierrevanhove/PicardFuchs/blob/main/PF-4sunset.ipynb}{4 loops},
\href{https://nbviewer.org/github/pierrevanhove/PicardFuchs/blob/main/PF-5sunset.ipynb}{5 loops}  analysed in  Section~\ref{sec:sunset}.
\item The differential operators for the multi-loop ice-cream cone graph at
  \href{https://nbviewer.org/github/pierrevanhove/PicardFuchs/blob/main/PF-triangle.ipynb}{1 loop},   \href{https://nbviewer.org/github/pierrevanhove/PicardFuchs/blob/main/PF-icecream-2loop.ipynb}{2 loops},
\href{https://nbviewer.org/github/pierrevanhove/PicardFuchs/blob/main/PF-icecream-3loop.ipynb}{3 loops}
analysed in Section~\ref{sec:icecream}
\item The differential operators for the \href{https://nbviewer.org/github/pierrevanhove/PicardFuchs/blob/main/PF-Kite.ipynb}{kite graph} of Section~\ref{sec:kite}
\item The differential operators for the
  \href{https://nbviewer.org/github/pierrevanhove/PicardFuchs/blob/main/PF-Tardigrade.ipynb}{tardigrade graph} of Section~\ref{sec:Tardigrade}.

\item The differential operators for the \href{https://nbviewer.org/github/pierrevanhove/PicardFuchs/blob/main/PF-DoubleBox.ipynb}{double-box graph}
  of Section~\ref{sec:DoubleBox}
\item The differential operators for the \href{https://nbviewer.org/github/pierrevanhove/PicardFuchs/blob/main/PF-Pentabox.ipynb}{pentabox graph}
  of Section~\ref{sec:Pentabox}

\end{itemize}

  \section{Feynman integrals} \label{sec:Feynman}

  In this section we review the connection between Feynman integrals and
relative period integrals.

\subsection{Definitions and notations}
The parametric representation of a Feynman integral associated to
Feynman graph $\Gamma$, with $n$ edges and $L$ (homology) loops,   is given by
the following projective integral in $\mathbb P^{n-1}(\mathbb R)$ in the variables
$(x_1,\dots,x_n)$
(see~\cite{Bloch:2005bh,Vanhove:2014wqa,Bitoun:2017nre})
\begin{equation}\label{e:GraphPara2}
    I_\Gamma(\underline s,\underline m^2;\underline \nu,D)=
    \int_{\Delta_n} \Omega_n(\underline s,\underline m^2;\underline \nu,D;\underline
    x),
\end{equation}
where
\begin{itemize}
\item The domain of integral is the simplex defined by the positive orthant
 \begin{equation}\label{e:DefDomain}
  \Delta_n:=\left\{[x_1,\dots,x_n]\in \mathbb P^{n-1}\ \middle\vert\  x_i\in\mathbb R,  x_i\geq0\right\}.
\end{equation}
\item The differential form is
  \begin{equation}
    \label{e:OmeganDef}
    \Omega_n(\underline s,\underline m^2;\underline \nu,D;\underline
    x):= \left(\mathcal U(\underline x)\over \mathcal U(\underline x)
      \mathcal L(\underline m^2,\underline x)-\mathcal V(\underline s,\underline x)\right)^\omega
    \, {\prod_{i=1}^n x_i^{\nu_i-1} \Omega^{(n)}_0\over \mathcal
      U(\underline x)^{D\over2}}
  \end{equation}
  and $\Omega^{(n)}_0$ is the natural differential $n-1$-form on the real projective space $\mathbb P^{n-1}$
\begin{equation}\label{e:Omega0def}
  \Omega^{(n)}_0:= \sum_{j=1}^n  (-1)^{j-1} \, x_j\, dx_1\wedge \cdots \wedge
  \widehat {dx_j}\wedge\cdots \wedge dx_n  \,,
\end{equation}
where $\widehat{dx_j}$ means that $dx_j$ is omitted in this sum.
\item The power is
  \begin{equation}\label{e:omegadef}
    \omega=\sum_{i=1}^n \nu_i-{LD\over2},
  \end{equation}
  with $L$ a
  positive integer. The powers $(\nu_1,\dots,\nu_n)$ are in $\mathbb Z^n$
  and $D$ is a real number. 
  \item A linear term with coefficients given by  the internal masses
  \begin{equation}
    \label{e:Ldef}
    \mathcal L(\underline m^2,\underline x):=\sum_{i=1}^n m_i^2 x_i+i\varepsilon
  \end{equation}
  with $\varepsilon$ a positive real number. 
  We denote by $\underline m^2:=(m_1^2,\dots,m_n^2)\in \mathbb C^n$
  the mass parameters. 
\end{itemize}
The  polynomials $\mathcal U(\underline x)$ and $\mathcal
V(\underline s,\underline
x)$ are determined by the Feynman graph $\Gamma$ using
graph theory, since the connection to  graph theory will not
be needed for the present work we refer
to~\cite{Nakanishi,Itzykson:1980rh,Bogner:2010kv,
  Weinzierl:2022eaz}. We only list the main properties of these polynomials.
\begin{itemize}
\item $\mathcal U(\underline x)$ is a homogeneous degree $L$
  polynomial in the variables $\underline x=(x_1,\dots,x_n)$ with
  coefficients in $\{0,1\}$.
  It is at most linear in each of the $x_i$ variables. This polynomial
  is determined by the  homology of the vacuum graph and 
  does not
  depend on the kinematic parameters $\underline s$ nor the masses
  $\underline m^2$~\cite{Nakanishi,Bogner:2010kv}.
\item $\mathcal V(\underline s,\underline x)$ is a homogeneous degree
  $L+1$ polynomial in the variables $x_1,\dots,x_n$ 
\begin{equation}\label{e:Fdef}
\mathcal V(\underline s,\underline x):=\sum_{i=1}^{N_v} s_i V_i(\underline x)
\end{equation}
where $V_{i}(\underline x)$ is are homogeneous polynomials
 of degree $L+1$   at most linear in each of
 the $x_i$ variables with
integer  coefficients. We denote by $\underline s:=(s_1,\dots,s_{N_v})$
 the kinematic variables.  The  number of
 independent kinematic variables 
 are constrained by the kinematic
 relations~\cite{Asribekov:1962tgp,Eden:1966dnq}. They imply
 relations between the coefficients of the graph polynomial $\mathcal
 V(\underline s,\underline x)$ and affect the singularity structure of
 integrand of the Feynman integral.

 \item The $i\varepsilon$ in~\eqref{e:Ldef} is the Feynman prescription
 for determining the contour of integration for the Feynman
 propagator. The integrals we will consider in this work exist in the limit
 $\varepsilon\to 0$ with $\varepsilon>0$, and the differential
 equations derived in this work are independent of the value of
 $\varepsilon$, and we will set $\varepsilon=0$ from now.  But for a proper 
 definition of the Feynman integrals and a correct determination of
 the solutions of the derived differential equation, one needs to
 reinstate the $\varepsilon$ dependence. We refer to~\cite{Hannesdottir:2022bmo} for a recent
 throughout discussion.
\end{itemize}

\subsection{Feynman integral as relative period integrals}

The integrand is a closed-form of the middle cohomology
\begin{equation}
\Omega_n(\underline s,\underline m^2;\nu,D;\underline x)   \in H^{n-1}(\mathbb P^{n-1}\backslash X_\Gamma),
\end{equation}
where $X_\Gamma$ is the polar part  of the Feynman integral
in~(\ref{e:GraphPara2}).  

When all the mass parameters are non-vanishing $m_i^2>0$ for $1\leq
i\leq n$, the Feynman integral is absolutely
convergent~\cite{Weinberg:1959nj}  when $\omega>0$ in~\eqref{e:omegadef}.
The conditions for the absolute convergence of the Feynman integrals
is a set of hyperplanes on the variables $(\underline \nu,D)$  in
$\mathbb C^{n+1}$~\cite{Speer:1975dc} .  
Because the polynomials $\mathcal U(\underline x)$, $\mathcal
L(\underline m^2,\underline x)$ and $\mathcal V(\underline
s,\underline x)$ are independent of the dimension of
space-time $D$ and the exponents $\underline \nu$, the Feynman integral is a meromorphic
function with singularities on linear hypersurfaces on  $(\underline \nu,D)$  in $\mathbb C^{n+1}$ as shown
by~\cite{Speer} and reproved by Panzer in~\cite{Panzer:2015ida}.

In the following, we will only consider the case of convergent
integrals with all non-vanishing masses, i.e. $m_i\neq0$ for $1\leq
i\leq n$ and $\omega>0$. Depending on the graph, it is still possible
to have some internal masses to vanish without introducing infrared
divergences and the Feynman integral is convergent. Such cases can be treated along the lines of the present
work. In these cases, the singular locus of the integrand will
change with the appearance of new singularities affecting the order of
the Picard--Fuchs operator.

Bloch, Esnault, and Kreimer showed that the
Feynman integral is a period integral of a mixed Hodge structure in
Eq.~(0.1) of~\cite{Bloch:2005bh}.
We would like to identify the algebraic
geometry behind the  question of
what kind of geometry is associated with the Feynman integral.
The Feynman integral is not a period of a smooth hypersurface (except for
some special cases like the two-loop sunset) and  singularity
structure of the integrand is rather
non-trivial (but this makes the Feynman integrals so rich).  One way to access a non-trivial information about
the rank of the Hodge structure is to perform a numerical evaluation
of the periods and their associated Picard--Fuchs differential equations.
From this information we can extract (an upper bound) on the dimension
of the algebraic variety for which the Feynman integral is period
integral. 

\subsection{Feynman integral differential equations}
The Feynman integrals are D-finite functions that satisfy finite order
inhomogeneous differential equations with respect to their physical
parameters~\cite{Laporta:2000dc,Smirnov:2010hn, Lee:2013hzt,
  Bitoun:2017nre}. 
For a fixed loop order $L$ and dimension $D$ the space defined by all
possible linear combinations of the Feynman integrals in~\eqref{e:GraphPara2}
\begin{equation}\label{eq:VGamma}
  V_\Gamma:= \sum_{\underline \nu\in\mathbb Z^n }  c_{\underline \nu}(D)I_\Gamma(\underline s,\underline m^2; \underline \nu,D) 
\end{equation}
is a finite dimensional vector space which dimension 
is given by the topological Euler characteristic of the complement of
the vanishing locus of the $\mathcal U$ and $\mathcal F=\mathcal
U\mathcal L-\mathcal V$ polynomials~\cite{Bitoun:2017nre}
\begin{equation}
  \dim(V_\Gamma)=(-1)^{n+1} \chi\left((\mathbb C^*)^n\backslash
    \mathbb V(\mathcal
      U)\cup \mathbb V(\mathcal F)\right),
  \end{equation}
  where
  \begin{align}
          \mathbb V(\mathcal
      U)&:=\{[x_1,\dots,x_n]\in\mathbb P^{n-1}\ \vert \ \mathcal U(\underline
      x)=0\};\cr \mathbb V(\mathcal F)&:=\{[x_1,\dots,x_n]\in\mathbb
                                        P^{n-1}\ \vert \ \mathcal U(\underline
      x)\mathcal L(\underline m^2,\underline x)-\mathcal V(\underline
      s,\underline x)=0\}.
  \end{align}
Since this vector space is finite-dimensional one can expand 
integral in the family of Feynman integrals $I_\Gamma(\underline s,\underline m^2; \underline
\nu,D)$ on a basis of, so-called master integrals,
$M_\Gamma(\underline s)$ with coefficients given by rational functions of
the parameters $\underline s$ ,$\underline m^2$, $\underline
\nu$ and the dimension $D$. The basis of master integral includes set
of indices $\underline\nu$ with possibly negative components. These
cases corresponds to integrals with irreducible scalar products of
loop momenta in the numerator in the momentum representation of the
Feynman integrals. 

Differentiating with respect to the physical parameters
$\underline s$, the master integrals $M_\Gamma(\underline s)$ satisfy the first order differential system of equations
(see~\cite{Henn:2014qga} for a physicist's review)
\begin{equation}\label{e:dM}
  dM_\Gamma(\underline s)= A_\Gamma\wedge M_\Gamma(\underline s)  .
\end{equation}
The matrix $A_\Gamma$ is a flat connection satisfying
\begin{equation}
  d  A_\Gamma+ A_\Gamma\wedge A_\Gamma=0.
\end{equation}
In positive integer dimension $D\in\mathbb N^*$ the connection $A_\Gamma$ is reducible and one
important question is to obtain the minimal order differential equation acting
on the Feynman integrals. This dimension in~\eqref{eq:VGamma}  gives
an upper bound on the order of the minimal order differential operator acting on the
Feynman integral.

Other approaches for deriving the system of differential operators
acting on Feynman integrals uses the GKZ approach~\cite{GKZ,Vanhove:2018mto,Klausen:2019hrg,Feng:2019bdx,Klemm:2019dbm,delaCruz:2019skx,Weinzierl:2022eaz}, for
constructing a D-module of differential operators. But this approach fails to lead to the complete set of
differential operators essentially because of the kinematic relations
between the monomials in the graph polynomials.  

The shortcomings of these approaches in providing directly a minimal
order differential equation for Feynman integrals in integer
dimensions, is the motivation of the present work.

\subsection{Relations between various integrals}
We consider the action of the differentiation with respect to the mass
parameters
\begin{equation}
 {\partial \over \partial m_i^2} \left(\mathcal U(\underline x)\over
   \mathcal U(\underline x)\mathcal L(\underline m^2,\underline
   x)-\mathcal V(\underline s,\underline x)\right)^\omega= -\omega\,  \left(\mathcal U(\underline x)\over
   \mathcal U(\underline x)\mathcal L(\underline m^2,\underline
   x)-\mathcal V(\underline s,\underline x)\right)^{\omega+1} x_i\,.
\end{equation}
We then conclude that the differentiating with respect to the mass
parameter $m_i^2$ shifts the value of the $\underline \nu$ by 
$\underline \nu_i=(0,\dots,0,1,0,\dots,0)$ where the $1$ is in
the $i$th position.
The anti-derivative (formal integration) with respect to a mass parameter 
\begin{equation}
\partial_{m_i^2}^{-1} \left(\mathcal U(\underline x)\over
   \mathcal U(\underline x)\mathcal L(\underline m^2,\underline
   x)-\mathcal V(\underline s,\underline x)\right)^\omega = {1\over\omega-1}\,  \left(\mathcal U(\underline x)\over
   \mathcal U(\underline x)\mathcal L(\underline m^2,\underline
   x)-\mathcal V(\underline s,\underline x)\right)^{\omega-1} x_i^{-1}\,.
\end{equation}
Therefore we can restrict ourself to the case where $\underline
\nu=(1,\dots,1)$.

If one considers the differential operator $\mathcal
U(\partial_{m_1^2},\dots,\partial_{m_n^2})$ where the variable $x_i$
is replaced by the partial derivative $\partial_{m_i^2}$ in the
$\mathcal U(\underline x)$ polynomial, we have
\begin{equation}
    \mathcal
U(\partial_{m_1^2},\dots,\partial_{m_n^2}) \Omega_n(\underline
s,\underline m;\underline \nu,D,\underline x)=\prod_{r=0}^{L-1}(-(\omega+r)) \Omega_n(\underline
s,\underline m;\underline \nu,D-2,\underline x)
\end{equation}
and the operator when $x_i$ are replaced by the 
$x_i^2\partial^{-1}_{m_i^2}$ we have
\begin{equation}
    \mathcal
U(x_1^2\partial^{-1}_{m_1^2},\dots,x_n^2\partial^{-1}_{m_n^2}) \Omega(\underline
s,\underline m;\underline \nu,D,\underline x)=\prod_{r=1}^L{1\over \omega-r} \Omega(\underline
s,\underline m;\underline \nu,D+2,\underline x).
\end{equation}
Such dimension shifting relations have been noticed by
Tarasov~\cite{Tarasov:1996br} using a different setup.

\section{ Picard--Fuchs equations for Feynman integrals }\label{sec:PFdef}
From now, we focus on the case where $\underline \nu=(1,\dots,1)$,
so that
\begin{equation}\label{e:OmegaGeneric}
  \Omega(t;D,\underline x):= \left(\mathcal
  U(\underline x)\over \mathcal U(\underline x)\mathcal L(\underline
  m^2,\underline x)- t\mathcal
  V(\underline s,\underline x)\right)^{n-{LD\over2}} {\Omega^{(n)}_0\over \mathcal
  U(\underline x)^{ {D\over2} } }   .
\end{equation}
We consider the case  of $D=2$ and $D=4$ dimensions
so that we have a rational differential
form in $\mathbb P^{n-1}(x_1,\dots,x_n)$, and convergent integral when
integrating this differential form over the positive orthant.
We have chosen to consider
the variation with respect to the parameter  $t$ as  an overall
scale
in front of  the kinematics graph polynomial  $\mathcal V(\underline
s,\underline x)$. One can consider a variation with respect to any
other physical parameter amongst the masses $m_i^2$ variable or any independent
kinematics $s_i$ variable. In that case, the denominator would take the
form $\mathcal F_0(\underline x)+t \mathcal F_1(\underline
x)$.
Clearly different choices for the parameters
$t$ amongst the kinematic or mass parameters, will lead to different
differential equations. But for generic values of the physical
parameters,  we conjecture that they all arise from the
same (singular) geometry determined by the singularity locus of the
integrand~\eqref{e:OmegaGeneric}.

We will derive the Picard--Fuchs operator differential operator
\begin{equation}
 \mathcal L_t=\sum_{r=0}^{o_i} q_r(\underline m,\underline s;t)\left(d\over dt\right)^r  
\end{equation}
such that $
\mathcal  L_t  \Omega(t;D,\underline x) = d\beta(\underline x,t)$,
for some holomorphic~$n-1$-form~$\beta(\underline x, t)$
on~$\mathbb{P}^{n-1}\setminus X_\Gamma$, corresponding to the
certificate in~\eqref{e:Cdef}.

Given any $n$-cycle $\gamma$ in~$\mathbb{P}^{n-1}\setminus X_\Gamma$, 
we obtain that
\begin{equation}\label{eq:1}
  \mathcal{L}_t  \int_\gamma \Omega(t;D,\underline x) = 0,
\end{equation}
so that~$\mathcal{L}_t$ is a differential equation satisfied by the
period integral defined by the integration of the differential form
over a cycle.

In the construction we will only consider the case where~$\beta(\underline x, t)$ is holomorphic on~$\mathbb{P}^{n-1}\setminus X_\Gamma$, that is
is $\beta(\underline x,t)$ does not have poles that are not present
in $\Omega(t;D,\underline x)$.
When~$\beta(\underline x, t)$ is only meromorphic, the left-hand side in~\eqref{eq:1} may not vanish after integration on~$\gamma$. This case may lead to an inhomogeneous differential equation
\begin{equation}
  \mathcal{L}_t \int_\gamma \Omega(t; D,\underline x) = \int_\gamma d \beta,
\end{equation}
where the right-hand side should reduce, with residue analysis, to an integral with one less variable, although the computation may not be straightforward.

To illustrate the difference between a holomorphic and a meromorphic term~$d \beta$,
consider the rational function
\begin{equation}
   F(x_1,x_2)=  { a x_1+b x_2+c  \over \left(\alpha x_1^2+\beta x_2^2+\gamma x_1
    x_2+\delta x_1+\eta x_2+\zeta\right)^2}
\end{equation}
where $a,b,c,\alpha,\beta,\gamma,\delta,\eta,\eta$ are independent of
$x_1$ and $x_2$ (this is the generic form of the multi-loop ice-cream
cone integrand
studied in Section~\ref{sec:icecream}.)
Using the creative telescoping algorithm~\cite{Koutchan,bostan2013creative} we find that
\begin{equation}\label{e:Fid}
 F(x_1,x_2)=\partial_{x_1}{ N_1(x_1,x_2)\over
   D_1(x_1,x_2)}+\partial_{x_2}{N_2(x_1,x_2)\over D_2(x_1,x_2)}
\end{equation}
where $N_i(x_1,x_2)$ with $i=1,2$ are polynomials in $x_1$ and $x_2$
and the denominator is (the coefficients are given on this page~\href{https://nbviewer.org/github/pierrevanhove/PicardFuchs/blob/main/Identity.ipynb}{identity})
\begin{align}
  \label{e:denD1D2}
  D_1(x_1,x_2)&=D_2(x_1,x_2)  \left(x_{2} (2 \alpha  b-a \gamma )-a \delta +2 \alpha  c\right),\cr
  D_2(x_1,x_2)&=2 \left(4 \alpha  \beta  \zeta -\alpha  \epsilon ^2-\beta  \delta ^2-\gamma ^2 \zeta
    +\gamma  \delta  \epsilon \right) \left(x_{2} (2 \alpha  b-a \gamma )-a \delta +2 \alpha  c\right) \cr
&\times  \left(\alpha x_1^2+\beta x_2^2+\gamma x_1
    x_2+\delta x_1+\eta x_2+\zeta\right)^2.
\end{align}
The denominator on the right-hand-side has a pole  at $x_{2}^0 =(a \delta -2 \alpha  c)/(2
\alpha  b-a \gamma )$ which is not a
pole of the original function. This means one can find a cycle where
$\gamma$ passing by $x_{2}^0$. Since $F(x_1,x_2)$ has no pole at $x_2=x_2^0$ the
integral of $\int_\gamma  F(x_1,x_2)$ is finite and
non-vanishing. Therefore,  the certificate must have extra poles that are not present in
 the original function.
Examples of this phenomenon were given by Picard~\cite{Picard1899} (see also~\cite{bostan2013creative}).

We will use the factorisation algorithm in~\cite{facto} to certify
whether the Picard--Fuchs operator is irreducible or not.
A differential operator~$\mathcal{L}$ is irreducible if it cannot be written as a composition
of two differential operators with order at least~1.
The minimal operator annihilating a given function need not be irreducible.
For example, the operator $t \frac{d^2}{dt^2} + \frac{d}{dt}$ is the minimal operator annihilating~$\log(t)$
and it factors as $\left( \frac{d}{dt} \right) \cdot \left( t \frac{d}{dt} \right)$.
However, an irreducible operator is the minimal annihilating operator of its non-zero solutions.
We encounter non-irreducible operators in the case of the Kite graph in four dimensions in Section~\ref{sec:Kite4D}.

\subsection{The Griffiths--Dwork reduction}\label{sec:algo}
The Griffiths--Dwork reduction is the main tool for computing Picard--Fuchs differential equations~\cite{Griffiths_1969,Dwork_1962,Dwork_1964},
although it needs to be completed by other reductions rules with the projective hypersurface~$X_\Gamma$ is singular.
This section describes briefly this reduction and its extension, following~\cite{Lairez}.

For a given~$D$, let us write~$\Omega(t;D, \underline x)$ as $\frac{A(\underline x) \Omega_0}{P(\underline x)^k}$,
where~$A$ and~$P$ are homogeneous polynomial in~$\underline x$ whose coefficients are rational functions in~$t$.
We assume moreover that~$P$ is square-free (as a polynomial in~$\underline x$).

It makes the computation easier to reformulate the integral in~$\mathbb{P}^{n-1}$ over an $n-1$-cycle $\gamma$
as an integral in~$\mathbb{C}^n$ over the $n$-cycle $\tilde\gamma = \left\{ \underline x\in \mathbb{C}^n \ \middle\vert\   [\underline x]\in \gamma \text{ and } \|\underline x\| = 1\right\}$:
\begin{equation}
  \int_\gamma \frac{A(\underline x)\Omega_0}{P(\underline x)^k} = \frac{1}{2\pi i} \int_{\tilde \gamma} \frac{A(\underline x)}{P(\underline x)^k} d\underline x,
\end{equation}
where~$d\underline x=dx_1\dotsb dx_n$.

Assume that~$A$ is in the Jacobian ideal of~$P$,
that is~$A = C_1 \frac{\partial P}{\partial x_1} + \dotsb  + C_n \frac{\partial P}{\partial x_n}$
for some homogeneous polynomials~$C_i$.
Then we can write
\begin{equation}
  Ad\underline x  = dP \wedge (C_1 \xi_1 + \dotsb + C_n \xi_n ),
\end{equation}
where~$\xi_i = (-1)^{i-1} dx_1 \dotsb \widehat{dx_i} \dotsb dx_n$.
Note that~$d \xi_i = dx_1\dotsb dx_n$, so
\begin{equation}\label{eq:4}
  (k-1) \frac{A}{P^k}d\underline x = \frac{\sum_i \frac{\partial C_i}{\partial x_i} }{P^{k-1}} d\underline x - d \left( \sum_i \frac{C_i}{P^{k-1}} \xi_i \right).
\end{equation}
The integral on~$\tilde \gamma$ of the exact differential vanishes, because we assume that~$P$ does not vanish on~$\tilde \gamma$, so, assuming~$k > 1$, we obtain that
\begin{equation}\label{eq:6}
   \int_{\tilde \gamma} \frac{A(\underline x)}{P(\underline x)^k} d\underline x = -\frac{1}{k-1} \int_{\tilde \gamma} \frac{\sum_i \frac{\partial C_i}{\partial x_i} }{P^{k-1}}d\underline x.
\end{equation}
The left-hand side is the integral of a form with pole order~$k$, while the right-hand side has pole order~$k-1$.
In the general case, when~$A$ is not in the Jacobian ideal, we may compute a normal form~$R(\underline x)$ of~$A$ modulo the Jacobian ideal (using for example a Gr\"obner basis)
that is
\begin{equation}
A = R + C_1 \frac{\partial P}{\partial x_1} + \dotsb  + C_n \frac{\partial P}{\partial x_n},\label{eq:2}
\end{equation}
with the property that~$R$ depends only on the class of~$A$ modulo the Jacobian ideal.
Then we have
\begin{equation}\label{eq:gdred}
   \int_{\tilde \gamma} \frac{A(\underline x)}{P(\underline x)^k} d \underline x =  \int_{\tilde \gamma} \frac{R(\underline x)}{P(\underline x)^k} d\underline x -\frac{1}{k-1} \int_{\tilde \gamma} \frac{\sum_i \frac{\partial C_i}{\partial x_i} }{P^{k-1}} d\underline x.
\end{equation}
The right-hand side is made of a term with a pole order~$k$ but reduced numerator,
and a term with pole order~$k-1$, to which we can apply the procedure recursively.
In the end, we obtain a decomposition
\begin{equation}\label{eq:3}
   \int_{\tilde \gamma} \frac{A(\underline x)}{P(\underline x)^k}d\underline x=  \sum_{i=1}^k \int_{\tilde \gamma} \frac{R_i(\underline x)}{P(\underline x)^i}d\underline x ,
\end{equation}
for some polynomials~$R_i(\underline x)$ in normal form with respect to the Jacobian ideal.
Moreover, $R_i$ is homogeneous of degree~$\deg R_i = i \deg P - n$.

\subsection{Picard--Fuchs equations in the smooth case}

In the case where the hypersurface~$X_\Gamma\subset \mathbb{P}^{n-1}$ is smooth,
the Jacobian ideal contains all the homogeneous polynomials of degree at least~$n (\deg P-1) - n + 1$ (this is the Macaulay bound)
this implies in particular that in the decomposition~\eqref{eq:3},
the polynomial~$R_i$ with~$i \geq n$ are all zero.
This gives the following algorithm to compute the Picard--Fuchs equation.

We compute the Griffiths--Dwork decomposition of
the $j$th derivative (with respect to~$t$) of the integral $I(t) = \int_{\tilde\gamma}  \frac{A(\underline x)}{P(\underline x)^k}d\underline x$
as
\begin{equation}
  \frac{d^j I}{d t^j} =
  \int_{\tilde \gamma}  \frac{A_j(\underline x)}{P(\underline x)^{k+j}}d\underline x=  \sum_{i=1}^{n-1} \int_{\tilde \gamma} \frac{R_{ji}(\underline x)}{P(\underline x)^i}d\underline x,
\end{equation}
where~$A_j$ is a homogeneous polynomial (of degree~$(k+i)\deg P - n$)
and the~$R_{ji}$ are homogeneous polynomials of degree~$i\deg P - n$.
The first equality is obtained by differentiating under the integral sign,
and the second equality is the Griffiths--Dwork reduction.

In view of the degree constraint on the polynomials~$R_{ji}$, the tuples $(R_{j1},\dotsc,R_{j,n-1})$,
for~$j\geq 0$, lie in a finite dimensional vector space
over the base field~$\mathbb{C}(t)$.
Given a~$N$ which exceeds this dimension, we can therefore compute coefficients~$a_i(t) \in \mathbb{C}(t)$,
not all zero,
such that
\begin{equation}
  a_N(t)  \frac{d^N I}{d t^N} + \dotsb + a_1(t) \frac{dI}{dt} + a_0(t) I(t) = 0,
\end{equation}
which is the desired differential equation,
simply by computing a non-vanishing solution of the linear system
\begin{equation}
  \sum_{j=0}^N a_j(t) R_{ji}(t; \underline x) = 0, \quad 1\leq i\leq n,
\end{equation}
with unknowns~$a_0,\dotsc,a_N$.

\subsection{An extension of the Griffiths--Dwork reduction}\label{sec:GDext}
When the hypersurface~$X_\Gamma \subset \mathbb{P}^{n-1}$ is not smooth, the
Griffiths--Dwork reduction step~\eqref{eq:gdred} may not be enough to reduce the
pole order when~$k \geq n$, independently on the numerator.
Other reduction rules come from the syzygies of the derivatives~$\frac{\partial P}{\partial x_i}$.
Let~$B_1,\dotsc,B_n$ be homogeneous of degree~$k \deg P - n+1$ such that~$\sum_i B_i \frac{\partial P}{\partial x_i} = 0$.
The tuple~$(B_1,\dotsc,B_n)$ is called a \emph{syzygy}.
The same computation as for~\eqref{eq:4} shows that
\begin{equation}\label{eq:5}
  \frac{\sum_i \frac{\partial B_i}{\partial x_i}}{P^k} d\underline x = d \left( \sum_i \frac{B_i}{P^{k}} \xi_i \right),
\end{equation}
so that
\begin{equation}
  \int_{\tilde \gamma} \frac{\sum_i \frac{\partial B_i}{\partial x_i}}{P^k} d\underline x = 0.
\end{equation}
In singular cases, these relations are missed by the Griffiths--Dwork reduction.

A syzygy $(B_1,\dotsc,B_n)$ is a \emph{trivial} syzygy if there are polynomials~$D_{ij}$
such that~$D_{ij}=-D_{ji}$ and
such that~$B_i = \sum_{j=1}^n D_{ij} \frac{\partial P}{\partial x_j}$,
(which implies immediately the relation~$\sum_i B_i \frac{\partial P}{\partial x_i} = 0$).
Trivial syzygies are irrelevant
because the corresponding relations are already reduced by the Griffiths--Dwork reduction.
Indeed, the numerator ${\sum_i \frac{\partial B_i}{\partial x_i}}$ in \eqref{eq:5} is in the Jacobian ideal:
\begin{equation}
 {\sum_i \frac{\partial B_i}{\partial x_i}} = \underbrace{\sum_i \sum_j D_{ij} \frac{\partial^2 P}{\partial x_j \partial x_i}}_{=0} + \sum_j\left(\sum_i \frac{\partial D_{ij}}{\partial x_i} \right) \frac{\partial P}{\partial x_j}.
\end{equation}

This leads to the following extension of the Griffiths--Dwork reduction.
Given a form $\frac{A}{P^k}\underline x$, with $A$ homogeneous of degree~$k\deg P -n$,
we first compute a basis of the space~$S_k$ of all syzygies of degree~$k\deg P-n+1$
quotiented by the space of trivial syzygies.
Then we compute a normal form~$R$ of~$A$ modulo the Jacobian ideal plus the space~$d V = \left\{ \sum_i\frac{\partial B_i}{\partial x_i}\ \middle\vert\ \underline B \in V \right\}$, that is
\begin{equation}
  A = R +  \underbrace{\sum_i\frac{\partial B_i}{\partial x_i}}_{\in dV} + \underbrace{C_1 \frac{\partial P}{\partial x_1} + \dotsb  + C_n \frac{\partial P}{\partial x_n}}_{\in \text{Jacobian ideal}},
\end{equation}
for some polynomials~$B_i$ and~$C_i$.
This leads to the following relation, similar to~\eqref{eq:4},
\begin{equation}\label{eq:4bis}
  (k-1) \frac{A}{P^k}d\underline x = \frac{\sum_i \frac{\partial C_i}{\partial x_i} }{P^{k-1}} d\underline x - d \left(  \sum_i \frac{B_i}{P^{k}} \xi_i + \sum_i \frac{C_i}{P^{k-1}} \xi_i \right).
\end{equation}
The integral on~$\tilde \gamma$ of the exact differential vanishes, so, assuming~$k > 1$, we obtain that
\begin{equation}
  \int_{\tilde \gamma} \frac{A(\underline x)}{P(\underline x)^k} d\underline x = -\frac{1}{k-1} \int_{\tilde \gamma} \frac{\sum_i \frac{\partial C_i}{\partial x_i} }{P^{k-1}}d\underline x,
\end{equation}
exactly similar to~\eqref{eq:6}.
We can apply the same reduction procedure to the right-hand side, by induction on the pole order.
Then we compute Picard--Fuchs differential equations in the same way as in the smooth case.

The extended Griffiths--Dwork reduction presented above is not always enough and
may need further extensions. There is a hierarchy of extensions which eventually
collapse to the strongest possible reduction. However, for all the computations
presented here, we only needed the first extension.

%%%%%%%%%%%%%%%%%%%%%%%%%%%%%%%%%%%%%%%%%%%%%%%%%%%%%%%%%%%%%%%%%%
\section{The multi-loop sunset graphs}\label{sec:sunset}

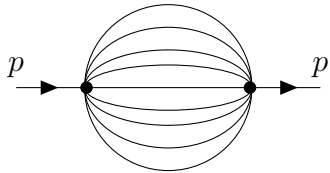
\begin{figure}[ht]
  \centering
      \begin{tikzpicture}[baseline=(x)]
        \begin{feynman}
       \vertex (x);
      \vertex[left=2cm of x,label=$p$](c1l);
      \vertex[right=2cm of x,label=$p$](c1r);
      \tikzfeynmanset{every vertex=dot}
        \vertex [left=1cm  of x] (xl);
        \vertex [right=1cm  of x] (xr);
        \diagram* {
         (c1l) -- [fermion] (xl);
         (xr) -- [fermion] (c1r);
         (xl) -- (xr);
				};
                              \end{feynman}
                                           \begin{pgfonlayer}{bg}
                                             \draw (x) circle (1.1cm);
                                              \draw (x) ellipse (1.1cm
                                              and .8cm);
                                                \draw (x) ellipse
                                                (1.1cm and .5cm);
                                                  \draw (x) ellipse (1.1cm and .3cm);
                        \end{pgfonlayer}
                   \end{tikzpicture}
  \caption{Multi-loop sunset}
  \label{fig:sunset}
\end{figure}

The multi-loop sunset graphs are  two-point graphs connected by $n$
edges as depicted in figure~\ref{fig:sunset}.
In two dimensions, the associated differential form in $\mathbb P^{n-1}(x_1,\dots,x_n)$  is given by
\begin{equation}
  \Omega^\su_n (t,\underline m^2) := {\Omega_0^{(n)}  \over\mathcal
  F^\su_n(t,\underline m^2;\underline x) }
\end{equation}
  with $\Omega_0^{(n)}$ the natural projective form on $\mathbb
  P^{n-1}$ defined in~\eqref{e:Omega0def} and  the degree $n$ homogeneous polynomial
  \begin{equation}
\label{e:FsunsetDef}
    \mathcal F^\su_n(p^2,\underline m^2;\underline x):= x_1\cdots x_n
    \left(\left(\sum_{i=1}^n {1\over x_i}\right)\left( \sum_{j=1}^n m_j^2x_j\right)-p^2\right) .
  \end{equation}
We  seek a minimal order  differential operator with respect to $t=p^2$
\begin{equation}
 \mathcal L_t= \sum_{r=0}^{o_n} q_r(t,\underline m^2) \left(d\over dt\right)^r   
\end{equation}
where $q_r(t,\underline m^2)$ is a polynomial in $t$ such that
\begin{equation}\label{e:PFsunset}
 \mathcal    L_t \Omega_n^\su (t,\underline m^2)=d\beta(t,\underline m^2)\,.
\end{equation}

For $n\geq4$ the Jacobian ideal  $J=\left\langle
\partial_{x_r}\mathcal F^\su_n(t,\underline m^2;\underline x), 1\leq r\leq n \right\rangle$ has non-isolated singularities at the positions
\begin{equation}
  x_i=x_j=0, \qquad \sum_{1\leq r\leq n\atop r\neq i, r\neq j}m_r^2 x_r=0.
\end{equation}
Therefore, we need to apply the Griffiths--Dwork reduction adapted to
this case as described in Section~\ref{sec:algo}.

 The equation $  \mathcal F^\su_n(t,\underline m^2;\underline x)$
  defines  families of Calabi--Yau manifolds associated to the
root lattice $A_n$ studied by Verrill~\cite{Verrill}. 
In this case, the denominator of period integral takes
the form of  $t-\phi_n(x_1,\dots,x_n)$ where $\phi_n(\underline x)$ is
a Laurent polynomial.  The period integrals associated to Laurent
polynomials are connected to mirror symmetry for Fano
manifolds~\cite{Batyrev:1998kx,Hori:2000kt,Coates},  and
our analysis gives supplementary support that the $n-1$-loop
sunset Feynman integrals are relative periods of
(singular) Calabi--Yau of dimension complex dimension
$n-2$~\cite{Bloch:2013tra,Bloch:2014qca,Bloch:2014qca,Bourjaily:2019hmc,DNV,Bonisch:2020qmm,Bonisch:2021yfw,Candelas:2021lkc,Forum:2022lpz}.

\begin{conj}[Order of the sunset Picard--Fuchs operator]\label{conjecture}
  The order of the minimal order Picard--Fuchs operator for generic
mass parameters is
\begin{equation}
o_n=2^{n}-\binom{n+1}{\left\lfloor \frac{n+1}{2}\right\rfloor }  ;
n\geq 2.
\end{equation}
The coefficient of the highest order derivative reads
\begin{equation}
  q_{o_n}(t,\underline m^2)= t^{\alpha(n)} \prod_{i=1}^{\vert\mu_n\vert} (t-\mu_n^i) \,
  \tilde q_{o_n}(t)   
\end{equation}
where $\alpha(n)\in\mathbb N$.  The polynomial $\tilde q_{o_n}(t)$ contains the
apparent singularities, and the non-apparent singularities are located
at $t=0$ and $t=\mu_n$ which are the $2^n$ so-called
thresholds
\begin{equation}\label{e:thresholds}
  \mu_n=\left\{\left(\sum_{i=1}^n \epsilon_i m_i\right)^2, (\epsilon_1,\dots,\epsilon_n)\in\{-1,1\}^n \right\}.
\end{equation}
where only distinct values of the squares are kept.
\end{conj}

\begin{itemize}
\item For $n=2$ this is shown in Section~\ref{sec:1sunset}
  \item For $n=3$ this is shown in Section~\ref{sec:2sunset}
  \item  For $n=4$ this is shown in Section~\ref{sec:3sunset}
  \item  For $n=5$ this is shown in Section~\ref{sec:4sunset}
    \item  For $n=6$ this is shown in Section~\ref{sec:5sunset}
        \item  For $n=7$ the algorithm gives an upper bound on the order
      of the differential operator matching the value $o_6$, see Section~\ref{sec:6sunset}.
\end{itemize}

    When all the mass parameters are identified $m_1^2=\cdots=m_n^2$ the
thresholds in~\eqref{e:thresholds} are located at the positions
\begin{equation}
  \mu_n=
  \begin{cases}
    \left\{0,2^2,\dots,n^2\right\}  & \text{for}~ n=0 \mod 2\cr
    \left\{1,3^2,\dots,n^2\right\}  & \text{for}~ n=1 \mod 2\cr    
  \end{cases}
\end{equation}
and the  minimal order differential operator in two-dimensions has of order $n-1$ and its form is
given by~\cite{Vanhove:2014wqa}
\begin{align}
  q_{n-1}(t)&=   t^{\left\lceil n\over2\right\rceil+{(-1)^{n-1}+1\over2}} \prod_{i=0}^{\lceil
          n\over2\rceil} (t-(n-2i)^2)\cr
              q_{n-2}(t)&=   {n-1\over2} {dq_{n-1}(t)\over dt}\cr
                          q_0(t)&=t-n
\end{align}
Generalisation to higher dimensions has been obtained in~\cite{MullerStach:2012mp,Klemm:2019dbm, Bonisch:2020qmm, Bonisch:2021yfw,Kreimer:2022fxm}.
In Section~\ref{sec:3sunset} and~\ref{sec:4sunset} we will show how the restriction of the masses leads to
the factorisation of the differential operator at three- and four-loop
orders respectively.

The certificate $Q$ in~\eqref{e:Q} or the $d\beta$
in~\eqref{e:PFsunset} leads to huge expressions that are evaluated by the algorithm,
since they assure that the telescoper annihilates the integrand but, we
will not consider them in this work. Our results are statements about
the Picard--Fuchs operator for the period integral  defined by the
integration over a cycle. In the
case of the sunset graph  the
integral over the torus has the large-$t$ series expansion (this is
the so-called maximal cut, discussed in~\cite{VanhoveISSAC})

\begin{equation}
  \pi_n^\su(t)=\int_{\vert x_1\vert =\cdots=\vert x_n\vert =1}
  \Omega^\su_n(t)=\sum_{m\geq0} t^{-m-1} \sum_{r_1+\cdots+r_m=m\atop
    r_i\geq0} \left( m!\over r_1!\cdots r_n!\right)^2  \,.
\end{equation}
The minimal order differential operator is this annihilator of this
series expansion $\mathcal L_t^n \pi_n^\su(t)=0$.

%------------------------------------------------------------------
\subsection{One-loop sunset}\label{sec:1sunset}

In this section, we recall the results for the Picard--Fuchs
operators  for the one-loop loop sunset integral
\begin{equation}
  \Omega_2^\su(t,\underline m^2)= {\Omega^{(2)}_0\over
    \mathcal F^\su_2(t,\underline m^2;\underline x)}   
\end{equation}
with
\begin{equation}
  \mathcal F^\su_2(t,\underline m^2;\underline x)  = t x_1 x_2-
    (m_1^2x_1+m_2^2x_2)\left({1\over x_1}+{1\over
        x_2}\right) x_1 x_2\,.
\end{equation}
The differential operator is readily obtained by applying the
Griffiths--Dwork reduction (see~\cite{Vanhove:2018mto}) to
get\footnote{The superscript denotes the number of independent mass
  parameter. When the mass parameters  are identified with $s$ different values, we use the
    notation $[r_1^{a_1} r_2^{a_2}\cdots r_s^{a_s}]$ such that
    $\sum_{i=1}^r a_i r_i=n$.
  When all
  the masses parameters are 
    different $m_1\neq \cdots \neq m_n$ we use $[1^n]$. } 
\begin{equation}
  \mathcal L_t^{[1^2]}= (t-(m_1+m_2)^2)(t-(m_1-m_2)^2) {d\over dt}+ t-m_1^2-m_2^2.
\end{equation}
The certificate has
\begin{equation}
  Q_1=2m_2^2x_2;  \qquad Q_2=-(m_1^2 + m_2^2 - p^2) x_2  
\end{equation}
so that
\begin{equation}
   \mathcal L_t^{[1^2]} {1\over 
    \mathcal F^\su_2(t,\underline m^2;\underline x)}   +\sum_{i=1}^2
  {\partial\over\partial x_i}   Q_i  {1\over 
    \mathcal F^\su_2(t,\underline m^2;\underline x)}  =0.
\end{equation}

%------------------------------------------------------------------
\subsection{Two-loop sunset}\label{sec:2sunset}

In this section, we recall the results for the Picard--Fuchs
operators  for the two-loop loop sunset integral
\begin{equation}
  \Omega_3^\su(t,\underline m^2)= {\Omega^{(3)}_0\over
    \mathcal F^\su_3(t,\underline m^2;\underline x)}   
\end{equation}
with
\begin{equation}
  \mathcal F^\su_3(t,\underline m^2;\underline x)  = t x_1\cdots x_3-
    (m_1^2x_1+\cdots+m_3^2x_3)\left({1\over x_1}+\cdots+{1\over
        x_3}\right) x_1\cdots x_3\,.
\end{equation}
In this case, the singular locus of the differential form defines a
smooth elliptic curve in $\mathbb P^2$
\begin{equation}
    \sum_{i=1}^3 x_i^{-1} \sum_{j=1}^3 m_j^2x_j-t=0.
\end{equation}
The Picard--Fuchs operator can be derived by applying the Griffiths--Dwork
reduction method~\cite{MullerStach:2011qkg,Bloch:2016izu,Vanhove:2018mto} giving the
second order operator
\begin{equation}
\mathcal L_t^{[1^3]}= \sum_{r=0}^2
q_r(t,\underline m^2) \left(d\over dt\right)^r .
\end{equation}
 The coefficients are polynomials in $t$ of degree $5+r$ with $0\leq
 r\leq 2$  are listed on this page \href{https://nbviewer.org/github/pierrevanhove/PicardFuchs/blob/main/PF-2sunset.ipynb}{2sunset}.  
The coefficient of the highest derivative term is given by
\begin{multline}
q_2(t,\underline m^2)=t \prod_{i=1}^4 (t-\mu_3^i)\cr\times ((m_1^2+m_2^2+m_3^2)^2-4(m_1^2m_2^2+m_1^2m_3^2+m_2^2m_3^2)+ 2 (m_1^2+m_2^2 + m_3^2)t - 3t^2)
\end{multline}
where we have introduced the thresholds contributions
\begin{equation}\label{e:thresholdsunset}
  \mu_3:=\{(-m_1+m_2+m_3)^2, (m_1-m_2+m_3)^2, (m_1+m_2-m_3)^2, (m_1+m_2+m_3)^2\}\,.  
\end{equation}
which are the position of the regular singularities. The roots of the
polynomial in the second line are apparent singularities.

%------------------------------------------------------------------
\subsection{Three-loop sunset}\label{sec:3sunset}
In this section, we give the results for the Picard--Fuchs
operators  for the three-loop sunset integral
\begin{equation}
  \Omega_4^\su(t,\underline m^2)= {\Omega^{(4)}_0\over
    \mathcal F_4(t,\underline m^2;\underline x)}   
\end{equation}
with
\begin{equation}
  \mathcal F^\su_4(t,\underline m^2;\underline x)  = t x_1\cdots x_4-
    (m_1^2x_1+\cdots+m_4^2x_4)\left({1\over x_1}+\cdots+{1\over
        x_4}\right) x_1\cdots x_4\,.
\end{equation}

We list the properties of the Picard--Fuchs operators for all the mass
configurations. The expressions are accessible online at~\href{https://nbviewer.org/github/pierrevanhove/PicardFuchs/blob/main/PF-3sunset.ipynb}{3sunset}

\begin{itemize}
  \item {\bf The four masses case $m_1\neq m_2\neq m_3\neq m_4$
    labelled $[1^4]$:}
When all the internal masses are different and all non-vanishing, the Picard--Fuchs operator is of
order 6
\begin{equation}
 \mathcal L_t^{[1^4]}= \sum_{r=0}^6 q_r^{[1^4]}(
t,\underline  m^2) \left(d\over dt\right)^r\,.
\end{equation}
the coefficients $q_r^{[1^4]}(t,\underline  m^2)$ are polynomials in $t$
of degree $21+r$ for $0\leq r\leq 6$.\footnote{For all the
  Picard--Fuchs operator considered in this work, the degree refers to
  the degree in $t$ of the polynomial multiplying the higher order
  derivative term. By homogeneity the degree of the polynomial
  coefficient  decrease with the derivative order. }
The coefficient of the highest order operator is

\begin{equation}
    q_6^{[1^4]}(t,\underline  m^2)=(t) \prod_{i=1}^8
    (t-(\mu^{(4)}_i)^2) \tilde q_6^{[1^4]}(t,\underline  m^2)
\end{equation}
with the thresholds contributions corresponding
\begin{multline}\label{e:threshold3sunset}
  \{\mu^{(4)}_i\}:=\big\{({m_1}+{m_2}+{m_3}+{m_4})^2,({m_1}+{m_2}+{m_3}-{m_4})^2,\cr({m_1}+{m_2}-{m_3}+{m_4})^2,
  ({m_1}+{m_2}-{m_3}-{m_4})^2,\cr
  ({m_1}-{m_2}+{m_3}+{m_4})^2,({m_1}-{m_2}+{m_3}-
  {m_4})^2,\cr ({m_1}-{m_2}-{m_3}+{m_4})^2,
  (-{m_1}+{m_2}+{m_3}+{m_4})^2\big\}
\end{multline}
and where  $ \tilde q_6^{[1^4]}(t,\underline  m^2)$ is a degree 17
polynomial in $t$.

The differential equation has only regular singularities located
at\footnote{These values correspond to  the position of the singular 
fibres of the pencil of $K3$ surfaces associated to the three-loop sunset (see the discussion
in~\cite{VanhoveStringMath2019} and~\cite{DNV})}
\begin{equation}\label{e:sing3sunset1}
  t=0,\infty,(\mu^{(4)}_i)^2 \qquad 1\leq i\leq 8\,, 
\end{equation}
and the apparent singularities at the roots of the degree 17 polynomial $\tilde q_6^{[1^4]}(t,\underline  m^2)$.

\item {\bf The three masses case $m_1\neq m_2\neq m_3=m_4$
    labelled $[1^22]$: }
  the Picard--Fuchs operator  has  order 5 
\begin{equation}
  \mathcal L^{[1^22]}_t= \sum_{r=0}^5 q_r^{[1^22]}(
t,\underline  m^2) \left(d\over dt\right)^r\,,
\end{equation}
the coefficients $q_r^{[1^22]}(t,\underline  m^2)$ are polynomials in $t$
of degree $12+r$ for $0\leq r\leq 5$.   
\item {\bf The two masses case $m_1=m_2\neq m_3=m_4$ labelled $[22]$:}
  The Picard--Fuchs operator of order 4
\begin{equation}
 \mathcal L^{[22]}_t= \sum_{r=0}^4 q_r^{[22]}(
t,\underline  m^2) \left(d\over dt\right)^r\,,
\end{equation}
the coefficients $q_r^{[22]}t,\underline  m^2)$ are polynomials in $t$
of degree $6+r$ for $0\leq r\leq 4$. 
\item {\bf The two masses case $m_1\neq m_2=m_3=m_4$ labelled $[13]$:}
  The Picard--Fuchs operator  of order 4
\begin{equation}
 \mathcal L^{[13]}_t= \sum_{r=0}^4 q_r^{[1^22]}(
t,\underline  m^2) \left(d\over dt\right)^r\,,
\end{equation}
the coefficients $q_r^{[13]}(t,\underline  m^2)$ are polynomials in $t$
of degree $5+r$ for $0\leq r\leq 4$. 
\item {\bf The single mass case $m_1=m_2=m_3=m_4$ labelled $[4]$:}
  The Picard--Fuchs operator  of  order 3 
\begin{equation}
 \mathcal L^{[4]}_t= \sum_{r=0}^5 q_r^{[3]}(
t,\underline  m^2) \left(d\over dt\right)^r\,.
\end{equation}
the coefficients $q_r^{[4]}(t,\underline  m^2)$ are polynomials in $t$
of degree $1+r$ for $0\leq r\leq3$. The Picard--Fuchs
operator was derived in~\cite{Verrill, Verrill3, Vanhove:2014wqa,Bloch:2014qca}.
It was shown in these references that this third order  Picard--Fuchs
operator is the symmetric square of the second order differential
operator
\begin{equation}
  \mathcal L_t^{[3]}= {t-8m_1^2\over 4t(t-4m_1^2)(t-16m_1^2)}
+{2(32m_1^2-15m_1t+t^2)\over
    t(t-4m_1^2)(t-16m_1^2)}   \left(d\over dt\right)   +\left(d\over dt\right)^2
\end{equation}
which is all equal mass Picard--Fuchs operator for the two-loop sunset
after the change of variables $t\to 4m_1^2-{(t-3m_1^2)^2\over t}$ and
the rescaling $f(t)\to \sqrt t f(t)$.
\end{itemize}

\subsubsection{Mass specialisation}
The Picard--Fuchs operator for the different mass configurations have
been derived using the extension of the Griffith-Dwork reduction
presented in Section~\ref{sec:GDext}.  

We analyse the relation between these different Picard--Fuchs operators.
When some masses are identified, the order of the minimal Picard--Fuchs operator
decreases.  Therefore by specialising the mass parameters, the Picard--Fuchs
operator becomes reducible as represented on this diagram
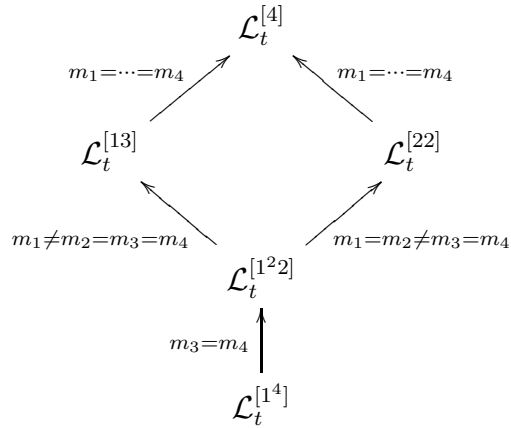
\begin{displaymath}
    \xymatrix{
        &\mathcal L_t^{[4]}\\
     \ar[ur]^{m_1=\cdots=m_4}   \mathcal L_t^{[13]}&&\mathcal L_t^{[22]}\ar[ul]_{m_1=\cdots=m_4}\\
        & \ar[ul]^{m_1\neq m_2=m_3=m_4}\mathcal L_t^{[1^22]}\ar[ur]_{m_1=m_2\neq m_3=m_4}\\
        &\mathcal L_t^{[1^4]}\ar[u]^{m_3=m_4}
}
\end{displaymath}
The arrows $P^a\to P^b$ represent the left factorisation of the  differential
operator  $P^a$, so that the operator $P^b$ divides the operator
$P^a$.  The factorisation can be obtained using  {\tt
    Maple}  {\tt DEtools} package or the factorisation algorithm~\cite{facto}. The
  page~\href{https://nbviewer.org/github/pierrevanhove/PicardFuchs/blob/main/PF-3sunset.ipynb}{3sunset} contains illustrative examples. 

We note that the certificate $d\beta$ factorises in a similar
way, and starting from the general differential operator for the four masses case, one
can derive all the special masses cases.

%------------------------------------------------------------------
\subsection{Four-loop sunset}\label{sec:4sunset}

In this section, we give the result for the  Picard--Fuchs
operators   for the four-loop  sunset integral

\begin{equation}
  \Omega_5^\su(t,\underline m^2)= {\Omega^{(5)}_0\over
    \mathcal F^\su_5(t,\underline m^2;\underline x)}   
\end{equation}
with
\begin{equation}\label{e:F4sunset}
  \mathcal F^\su_5(t,\underline m^2;\underline x)  = t x_1\cdots x_5-
    (m_1^2x_1+\cdots+m_5^2x_5)\left({1\over x_1}+\cdots+{1\over
        x_5}\right) x_1\cdots x_5\,.
\end{equation}y

We list the properties of the Picard--Fuchs operators for all the mass
configurations. The expressions for the coefficients are accessible online~\href{https://nbviewer.org/github/pierrevanhove/PicardFuchs/blob/main/PF-4sunset.ipynb}{4sunset}.

\begin{itemize}
  \item {\bf The five  masses are different $m_1\neq m_2\neq
m_3\neq m_4\neq m_5$ and all non-vanishing configuration:} the Picard--Fuchs operator is of
order 12
\begin{equation}
 \mathcal L^{[1^5]}_t= \sum_{r=0}^{12} q_r^{[1^5]}(
t,\underline  m^2) \left(d\over dt\right)^r\,.
\end{equation}
the coefficients $q_r^{[1^5]}(t,\underline  m^2)$ are polynomials in $t$
of degree $109+r$ for $0\leq r\leq 12$.
The coefficient $q_{12}^{[1^5]}(t,\underline  m^2)$ has the
following form
\begin{equation}
    q_{12}^{[1^5]}(t,\underline  m^2)=(t)^{12}\prod_{i=1}^{16}
    (t-\mu_i^2)  \tilde q_{12}^{[1^5]}(t,\underline  m^2)
  \end{equation}
  where $\tilde q_{12}^{[1^5]}(t,\underline  m^2)$ is a degree 98
  polynomial in $t$, and the $\mu_i$ are the 16 thresholds
  \begin{multline}
    \{\mu_i\}:=\big\{({m_1}+{m_2}+{m_3}+{m_4}+{m_5})^2,({m_1}+{m_2}+{m_3}+{m_4}-{m_5})^2,\cr
    ({m_1}+{m_2}+{m_3}-{m_4}+{m_5})^2,({m_1}+{m_2}+{m_3}-{m_4}-{m_5})^2,({m_1}+{m_2}-{m_3}+{m_4}+{m_5})^2,\cr
    ({m_1}+{m_2}-{m_3}+{m_4}-{m_5})^2,({m_1}+{m_2}-{m_3}-{m_4}+{m_5})^2,(-{m_1}-{m_2}+{m_3}+{m_4}+{m_5})^2,\cr
    ({m_1}-{m_2}+{m_3}+{m_4}+{m_5})^2,({m_1}-{m_2
   }+{m_3}+{m_4}-{m_5})^2,({m_1}-{m_2}+{m_3}-{m_4}+{m_5}
   )^2,\cr
   (-{m_1}+{m_2}-{m_3}+{m_4}+{m_5})^2,({m_1}-{m_2}-{m_3}+{m_4}+{m_5})^2,(-{m_1}+{m_2}+{m_3}-{m_4}+{m_5})^2,\cr
   (-
   {m_1}+{m_2}+{m_3}+{m_4}-{m_5})^2,(-{m_1}+{m_2}+{m_3}+
   {m_4}+{m_5})^2\big\}    
  \end{multline}

The differential equation has only regular singularities
located\footnote{They correspond to the positions of singular 
fibres of the pencil of Calabi--Yau three-fold (see the discussion
in~\cite{VanhoveStringMath2019} and~\cite{DNV}).}
\begin{equation}\label{e:sing4sunset1}
  t=0,\infty, \mu_i \qquad 1\leq i\leq 16\,,
\end{equation}
and the apparent singularities at the roots of the degree 98
polynomial $\tilde q_{12}^{[1^5]}(t,\underline  m^2)$.

\item {\bf The four  different mass configuration} $m_1^2\neq m_2^2\neq m_3^2\neq m_4^2=m_5$ The differential
  operator has order 10. 
 The degree of the
polynomial coefficient $q^{[1^41]}_r(t,\underline  m^2)$ are $57+r$ with
$0\leq r\leq 10$. 
\item {\bf The three  different mass configuration:}  the differential
operator has order 8.
  \begin{itemize}
    \item 
    $m_1\neq m_2\neq m_3=m_4=m_5$: the degree of the
polynomial coefficient $q^{[1^23]}_r(t,\underline  m^2)$ are $26+r$ with
$0\leq r\leq 8$.
   \item  
     $m_1\neq m_2=m_3\neq m_4=m_5$: the degree of the
polynomial coefficient $q^{[12^2]}_r(t,\underline  m^2)$ are $33+r$ with
$0\leq r\leq 8$. 
\end{itemize}
\item  {\bf The two  different mass configuration:} the differential
operator has order 6.
  \begin{itemize}
\item          $m_1\neq m_2=m_3=m_4=m_5$: the degree of the
polynomial coefficient $q^{[14]}_r(t,\underline  m^2)$ are $8+r$ with
$0\leq r\leq 6$. 
\item            $m_1=m_2\neq m_3=m_4=m_5$:  the degree of the
polynomial coefficient $q^{[23]}_r(t,\underline  m^2)$ are $13+r$ with
$0\leq r\leq 6$.
\end{itemize}
            \item  {\bf The all equal mass configuration
$m_1=m_2=m_3=m_4=m_5$:} the differential
operator  has order 4. The degree of the
polynomial coefficient $q^{[5]}_r(t,\underline  m^2)$ are $1+r$ with
$0\leq r\leq 4$.
This case has been derived in~\cite[\S9]{Vanhove:2014wqa}.
\end{itemize}

\subsubsection{Mass specialisation}
The Picard--Fuchs operator for the different mass configurations have
been derived using the extension of the Griffith-Dwork reduction
presented in Section~\ref{sec:GDext}.  

We analyse the relation between these different Picard--Fuchs operators.
When some masses are identified, the  order of the minimal Picard--Fuchs operator
decreases.  By specialising the mass parameters, the Picard--Fuchs
operator becomes reducible as represented on this diagram

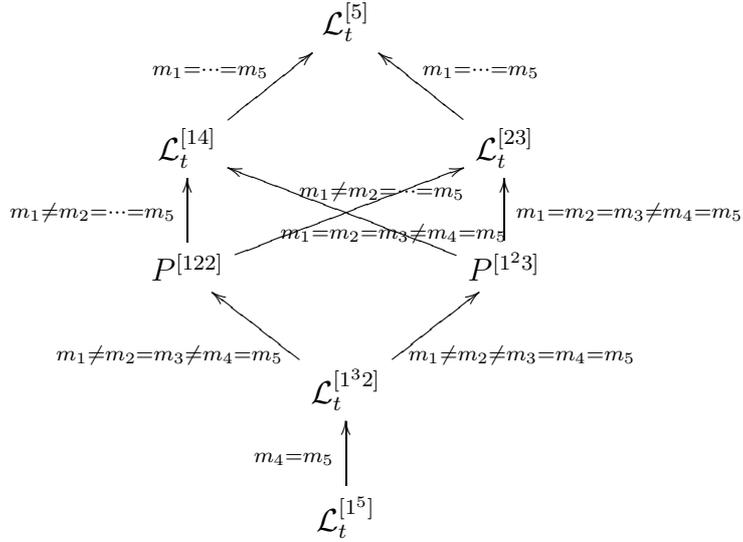
\begin{displaymath}
  \xymatrix{
    &\mathcal L_t^{[5]}\\
     \ar[ur]^{m_1=\cdots=m_5}   \mathcal L_t^{[14]}&&\mathcal L_t^{[23]}\ar[ul]_{m_1=\cdots=m_5}\\
\ar[u]^{m_1\neq m_2=\cdots=m_5}     \ar[urr]_{m_1=m_2=m_3\neq m_4=m_5}
P^{[122]}&&
P^{[1^23]}\ar[ull]_{m_1\neq m_2=\cdots=m_5}\ar[u]_{m_1=m_2=m_3\neq m_4=m_5}
\\
        & \ar[ul]^{m_1\neq m_2=m_3\neq m_4=m_5}
       \mathcal L_t^{[1^32]}\ar[ur]_{m_1\neq m_2\neq m_3=m_4=m_5}\\
       &\mathcal L_t^{[1^5]}\ar[u]^{m_4=m_5}
}
\end{displaymath}
The arrows $P^a\to P^b$ represent the left factorisation of the  differential
operator  $P^a$, so that the operator $P^b$ divides the operator
$P^a$ as can be checked using the {\tt SageMath} {\tt ore\_algebra} package~\cite{ore,ore2}.

The difference here with the three-loop sunset case of
Section~\ref{sec:3sunset} is that the order Picard--Fuchs operators
decreases by 2 when two masses
parameters are identified. It will be explained in~\cite{DNV}  that this is a consequence of the changes in the
cohomology for 
Calabi--Yau three-fold geometry determined by the sunset graph polynomial~\eqref{e:F4sunset}.

%------------------------------------------------------------------
\subsection{The five-loop sunset}\label{sec:5sunset}

In this section, we give the result for the  Picard--Fuchs
operators   for the five-loop sunset integral

\begin{equation}
  \Omega_6^\su(t,\underline m^2)= {\Omega^{(6)}_0\over
    \mathcal F^\su_6(t,\underline m^2;\underline x)}   
\end{equation}
with
\begin{equation}
  \mathcal F^\su_6(t,\underline m^2;\underline x)  = t x_1\cdots x_6-
    (m_1^2x_1+\cdots+m_6^2x_6)\left({1\over x_1}+\cdots+{1\over
        x_6}\right) x_1\cdots x_6\,.
\end{equation}
We list the properties of the Picard--Fuchs operators for all the mass
configurations. Numerical results are given on this page \href{https://nbviewer.org/github/pierrevanhove/PicardFuchs/blob/main/PF-5sunset.ipynb}{5sunset}.
\begin{itemize}
\item {\bf The six  mass configuration $m_1\neq m_2\neq
m_3\neq m_4\neq m_5\neq m_6$ denoted $[1^6]$:} the
Picard--Fuchs operator  of order 29 and degree of the
polynomial $q_{29}(t)$ is 521.
\item {\bf The five  mass configuration  $m_1\neq m_2\neq
m_3\neq m_4\neq m_5= m_6$ denoted $[1^42]$:} the Picard--Fuchs operator is of
order 23 and degree of the
polynomial $q_{23}(t)$ is 305.
\item {\bf The four  mass configuration $[1^33]$ and $[1^22^2]$:}
  the Picard--Fuchs operator $\mathcal L_t^{[1^33]}$ is of
order 17 and degree of the
polynomial $q^{[1^33]}_{17}(t)$ is 142, and the Picard--Fuchs operator
$\mathcal L_t^{[1^22]}$ is of
order 18 and degree of the
polynomial $q^{[1^22]}_{18}(t)$ is 174.
\item {\bf The three  mass configuration $[1^24]$,   $[123]$ and
    $[2^3]$:} the Picard--Fuchs operator $\mathcal L_t^{[1^24]}$ is of
order 12 and degree of the
polynomial $q^{[1^24]}_{12}(t)$ is 57. The Picard--Fuchs operator
$\mathcal L_t^{[123]}$ is of
order 13 and degree of the
polynomial $q^{[123]}_{13}(t)$ is 79.
\item {\bf The two  mass configuration $[15]$, $[24]$, $[3^2]$:} the
  Picard--Fuchs operator $\mathcal L_t^{[15]}$ is of
order 8 and degree of the
polynomial $q^{[15]}_{8}(t)$ is 20. The
  Picard--Fuchs operator $L_t^{[24]}$ is of
order 9 and degree of the
polynomial $q^{[24]}_{9}(t)$ is 31. The
  Picard--Fuchs operator $L_t^{[3^2]}$ is of
order 9 and degree of the
polynomial $q^{[3^2]}_{9}(t)$ is 34
\item {\bf The one  mass case $m_1= \cdots= m_6$ $[6]$:} the Picard--Fuchs operator is of
order 5, the degree of the polynomial $q_5(t)$ is 17. The operator is given in~\cite{Vanhove:2014wqa}.
\end{itemize}

\subsection{The six-loop sunset}\label{sec:6sunset}
In this section, we give the result for the  Picard--Fuchs
operators   for the six-loop  sunset integral

\begin{equation}
  \Omega_7^\su(t,\underline m^2)= {\Omega^{(7)}_0\over
    \mathcal F^\su_7(t,\underline m^2;\underline x)}   
\end{equation}
with
\begin{equation}
  \mathcal F^\su_7(t,\underline m^2;\underline x)  = t x_1\cdots x_7-
    (m_1^2x_1+\cdots+m_7^2x_7)\left({1\over x_1}+\cdots+{1\over
        x_7}\right) x_1\cdots x_7\,.
  \end{equation}
  In this case, the full computation of the Picard--Fuchs operator was not
  possible. We could only compute it for fixed random integer masses, performing all arithmetic operations
  modulo a large prime. The algorithm gave an order of 58  with a
  degree 2273
  for the head polynomial. The full computation would have lasted for several weeks.
  The order is compatible with the
  conjecture~\ref{conjecture}.

%%%%%%%%%%%%%%%%%%%%%%%%%%%%%%%%%%%%%%%%%%%%%%%%%%%%%%%%%%%%%%%%%%
\section{The multi-scoop ice-cream cone graphs}\label{sec:icecream}

\begin{figure}[ht]
  \centering
      \begin{tikzpicture}[baseline=(x)]
        \begin{feynman}
       \vertex (x);
      \vertex[left=2cm of x,label=$p_1$](c1l);
      \vertex[right=2cm of x,label=$p_3$](c1r);
      \vertex[below=3cm of x,label=left:$p_2$,](c1b);
      \tikzfeynmanset{every vertex=dot}
        \vertex [left=1cm  of x] (xl);
        \vertex [right=1cm  of x] (xr);
          \vertex [below=2cm  of x] (xb);
        \diagram* {
         (c1l) -- [fermion] (xl);
         (c1r) -- [fermion] (xr);
          (c1b) -- [fermion] (xb);
         (xl) -- (xr);
         (xl) -- (xb);
          (xb) -- (xr);
                                };
                              \end{feynman}
                                           \begin{pgfonlayer}{bg}
                                                \draw (x) ellipse
                                                (1.1cm and .5cm);
                                                  \draw (x) ellipse (1.1cm and .3cm);
                        \end{pgfonlayer}
                   \end{tikzpicture}
  \caption{Multi-loop ice-cream cone with $n$ loops has  $n-1$ scoops.}
  \label{fig:icecream}
\end{figure}
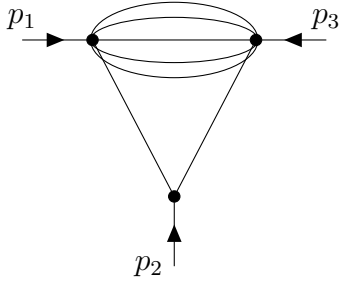

The multi-scoop ice-cream cone graphs represented in
fig.~\ref{fig:icecream} are 
three-point graphs, with $p_1+p_2+p_3=0$,   obtained
by splitting one edge of the multi-loop sunset graphs.

 The differential form for the $n$-scoop ice-cream cone in two
 dimension in $\mathbb P^{n+2}(x_1,\dots,x_{n+3})$ reads
 \begin{equation}\label{e:OmegaIceCream}
   \Omega_n(t,\underline m^2,\underline p^2)=   {\mathcal U_n(\underline x)\over \left(\mathcal
       U_n(\underline x) \mathcal L_n(\underline m^2,\underline
       x)-t\mathcal V_n(\underline m^2,\underline p^2;\underline x)\right)^2} \Omega_0^{(n+2)}
 \end{equation}
 with the following graph polynomials
 \begin{align}\label{e:PolIceCream}
   \mathcal U_n(\underline x)&:=\left((x_1+x_2)\left(\sum_{i=3}^{n+3}
                               {1\over x_i}\right) +1 \right)\prod_{i=3}^{n+3} x_i\cr
                             \mathcal L_n(\underline m^2,\underline
                             x)&=m_1^2x_1+\cdots+m_{n+2}^2 x_{n+2}\cr
\mathcal V_n(\underline p^2,\underline x)&=\left(
                                           \left(p_1^2x_1+p_3^2x_2\right)+
                                           p_2^2x_1x_2
                                           \left(\sum_{i=3}^{n+3} {1\over x_i}\right)
                                           \right)\prod_{i=3}^{n+3}x_i.
 \end{align}
We are considering the case where all the momenta are scaled by
$\sqrt{t}$, i.e. $(p_1,p_2,p_3)\to \sqrt{t}(p_1,p_2,p_3)$,  henceforth
the $t$ factor multiplying $\mathcal V_n(\underline m^2,\underline
p^2;\underline x)$ in eq.~\eqref{e:OmegaIceCream}.
 
The ice-cream cone graph Jacobian ideal 
\begin{equation}
  J_n:=\left\langle \partial_z \left(\mathcal
       U_n(\underline x) \mathcal L_n(\underline m^2,\underline
       x)-t\mathcal V_n(\underline m^2,\underline p^2;\underline
       x)\right), z\in \{x_1,\dots,x_{n+3}\}\right\rangle   
\end{equation}
vanishes for
\begin{equation}
  x_{i+3}=x_{j+3}=0, \quad  t p_2^2 x_1 x_2 + (x_1 + x_2) \left(\sum_{r=0\atop
    r\neq i,j}^n m_{3+r}^2x_{3+r}\right)=0,\quad  0\leq i,j\leq n.
\end{equation}

We will determine  the Picard--Fuchs operator with respect to the
parameter $t$
\begin{equation}
 \mathcal  L_t^n=\sum_{r=0}^{o_n} q_r^n(  t) \left(d\over dt\right)^r
\end{equation}
acting on 
the multi-scoop ice-cream cone  as
\begin{equation}
\mathcal  L_t^n\left(    {\mathcal U_n(\underline x)\over \left(\mathcal
       U_n(\underline x) \mathcal L_n(\underline m^2,\underline
       x)-t\mathcal V_n(\underline m^2,\underline p^2;\underline
       x)\right)^2}\right)=\sum_{i=1}^{n+3} \partial_{x_i} Q_i(\underline
 x,t)
\end{equation}
The
certificates $Q_i$ and $R_i$ must not have poles that are not present in the
original as discussed in Section~\ref{sec:PFdef}.

The results for Picard--Fuchs
operator for  the zero-scoop
ice-cream cone graph (the triangle graph) in Section~\ref{sec:0scoop},
for the
one-scoop ice-cream cone graph in Section~\ref{sec:1scoop} 
and finally for 
the two-scoop ice-cream cone graph in Section~\ref{sec:2scoop}.

\subsection{The zero-scoop ice-cream cone (triangle) graph}\label{sec:0scoop}

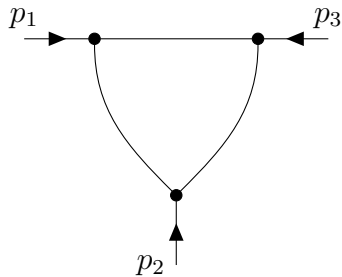
\begin{figure}[ht]
  \centering
   \begin{tikzpicture}[baseline=(x)] 
          \begin{feynman}
       \vertex (x);
      \vertex[left=2cm of x,label=$p_1$](c1l);
      \vertex[right=2cm of x,label=$p_3$](c1r);
      \vertex[below=3cm of x,label=left:$p_2$,](c1b);
      \tikzfeynmanset{every vertex=dot}
        \vertex [left=1cm  of x] (xl);
        \vertex [right=1cm  of x] (xr);
          \vertex [below=2cm  of x] (xb);
        \diagram* {
         (c1l) -- [fermion] (xl);
         (c1r) -- [fermion] (xr);
          (c1b) -- [fermion] (xb);
         (xl) -- (xr);
         (xb) --[out=135,in=-90]  (xl);
          (xb) --[out=45,in=-90] (xr);
				};
                              \end{feynman}     
                \end{tikzpicture}
                \caption{The zero-scoop (one-loop) ice-cream cone graph
                                \label{fig:icecream0scoop}
}
\end{figure}

In this section, we give the Picard--Fuchs operator for
the zero-scoop (one-loop) ice-cream cone graph of
figure~\ref{fig:icecream0scoop} corresponding to  the case  $n=0$
in~\eqref{e:OmegaIceCream} and~\eqref{e:PolIceCream}. The results are
summarised on the page~\href{https://nbviewer.org/github/pierrevanhove/PicardFuchs/blob/main/PF-triangle.ipynb}{PF-triangle}

With this  case we illustrate how the regularity of the certificate
affects the form of the differential equation.
For the single-scale triangle graph we have for the special
configuration $m_1^2=m_2^2=m_3^2=p_2^2=p_3^2=1$ and $p_1^2=5$ in the
patch $x_3=1$
\begin{multline}
 \frac{x_{1}+x_{2}+1}{(t (x_{1} (x_{2}-5)+x_{2})-(x_{1}+x_{2}+1)^2)^2}\cr
=\partial_{x_1}\left(\frac{(x_{2}-7) (x_{1}+x_{2}+1) \left((4 x_{1}+9) x_{2}^2-5 (7 x_{1}+6)
   x_{2}+85 x_{1}-15\right)}{2 (t-9) (x_{2}-5)^2 (t (x_{1} (x_{2}-5)+x_{2})-(x_{1}+x_{2}+1)^2)^2}\right)\cr
+\partial_{x_1}\left(\frac{-t(x_{2}-7) \left(x_{1}^2 (x_{2}-5)^2+x_{1} ((x_{2}-8)
   x_{2}+35) (x_{2}-5)\right)}{2 (t-9) (x_{2}-5)^2 (t (x_{1} (x_{2}-5)+x_{2})-(x_{1}+x_{2}+1)^2)^2}\right)\cr
+\partial_{x_1}\left(\frac{-t (x_{2}-7) \left(x_{2} ((x_{2}-4) x_{2}+10)-25\right)}{2 (t-9) (x_{2}-5)^2 (t (x_{1} (x_{2}-5)+x_{2})-(x_{1}+x_{2}+1)^2)^2}\right)\cr
+\partial_{x_2}\left(\frac{-\left((x_{2}-7) \left(t (x_{2}-5)^2-4 \left(x_{2}^2-5
   x_{2}-5\right)\right) (x_{1}+x_{2}+1)\right)}{2 (t-9) (x_{2}-5) \left(t (x_{1} (x_{2}-5)+x_{2})-(x_{1}+x_{2}+1)^2\right)^2}\right).
\end{multline}
The certificate has a pole at $x_2=5$ which is not a pole of the
original integral. Therefore,  integrating over a cycle passing through
that pole will not be possible. Since the Feynman integral is defined
as integrated over the positive orthant~\eqref{e:DefDomain}, we
cannot allow this certificate. 

For a regular certificate, the Picard--Fuchs operator is order 1 given by

\begin{equation}
 \mathcal L_t^1=q_0(t) + q_1(t) \,{d\over dt}
\end{equation}
with
\begin{multline}
  q_1(t)=-\left(t p_2^2-(m_1+m_2)^2\right) \left(t p_2^2-(m_1-m_2)^2\right) \cr
  \times\left(m_{1}^2 (p_1^2-p_2^2-p_3^2)-m_{2}^2
   (p_1^2+p_2^2-p_3^2)+2 m_3^2 p_2^2-p_2^2 
   (p_1^2-p_2^2+p_3^2)t\right) \cr
 \times\big(p_1^2p_2^2p_3^2t^2+\left(-\left(m_{1}^2 p_3^2 (p_1^2+p_2^2-p_3^2)\right)+m_{2}^2
   p_1^2 (p_1^2-p_2^2-p_3^2)-m_3^2 p_2^2
   (p_1^2-p_2^2+p_3^2)\right) t\cr
 +m_{1}^4 p_3^2-m_{1}^2 \left(m_{2}^2
   (p_1^2-p_2^2+p_3^2)+m_3^2
   (-p_1^2+p_2^2+p_3^2)\right)+m_{2}^4 p_1^2-m_{2}^2
   m_3^2 (p_1^2+p_2^2-p_3^2)+m_3^4 p_2^2\big).
 \end{multline}
 The expression for 
$q_0(t)$ is too large to be displayed here but it is given on this page
\href{https://nbviewer.org/github/pierrevanhove/PicardFuchs/blob/main/PF-triangle.ipynb}{PFtriangle}.
\subsection{The one-scoop ice-cream cone graph}
\label{sec:1scoop}

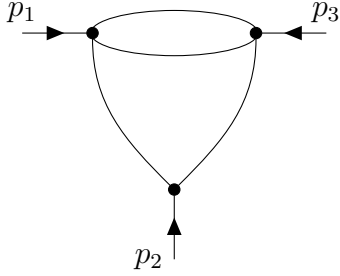
\begin{figure}[ht]
  \centering
    \begin{tikzpicture}[baseline=(x)]
        \begin{feynman}
       \vertex (x);
      \vertex[left=2cm of x,label=$p_1$](c1l);
      \vertex[right=2cm of x,label=$p_3$](c1r);
      \vertex[below=3cm of x,label=left:$p_2$,](c1b);
      \tikzfeynmanset{every vertex=dot}
        \vertex [left=1cm  of x] (xl);
        \vertex [right=1cm  of x] (xr);
          \vertex [below=2cm  of x] (xb);
        \diagram* {
         (c1l) -- [fermion] (xl);
         (c1r) -- [fermion] (xr);
          (c1b) -- [fermion] (xb);
       %  (xl) -- (xr);
         (xb) --[out=135,in=-90]  (xl);
          (xb) --[out=45,in=-90] (xr);
				};
                              \end{feynman}
                                           \begin{pgfonlayer}{bg}
                                             %   \draw (x) ellipse                                                (1.1cm and .5cm);
                                                  \draw (x) ellipse (1.1cm and .3cm);
                        \end{pgfonlayer}
                      \end{tikzpicture}
                                              \caption{The one-scoop (two-loop) ice-cream
cone                            graph
                          \label{fig:icecream1scoop}}
\end{figure}
In this section, we give the Picard--Fuchs operator for 
the one-scoop (two-loop) ice-cream cone graph of
figure~\ref{fig:icecream1scoop} corresponding to  the case $n=1$  in~\eqref{e:OmegaIceCream} and~\eqref{e:PolIceCream}.\footnote{This
  is sometimes called the Dunce's cap graph e.g.~\cite{Klausen:2021yrt}. But since we are generalising to
the multi-loop case we will call these graphs ice-cream cone with
multi-scoops.} The results are summarised on the
page~\href{https://nbviewer.org/github/pierrevanhove/PicardFuchs/blob/main/PF-icecream-2loop.ipynb}{PF-icecream-2loop}.

 With a regular certificate, the Picard--Fuchs operator is of order 2 and degree 9
 \begin{equation}
  \mathcal L_t^2=q_0(t)+q_1(t) {d\over dt}+q_2(t) \left(d\over dt\right)^2, 
 \end{equation}
 the higher order coefficient is given by
 \begin{equation}
   q_2(t)=(tp_2^2-(m_1+m_2)^2)(tp_2^2-(m_1-m_2)^2) c_1(t) c_2(t)c_3(t)
 \end{equation}
 with
 \begin{multline}
   c_1(t)=   p_{1}^2 p_{2}^2 p_{3}^2 t^2 +t\Big(m_{1}^2 \left(-p_{1}^2 p_{3}^2-p_{2}^2 p_{3}^2+p_{3}^4\right)+m_{2}^2 \left(p_{1}^4-p_{1}^2
   p_{2}^2-p_{1}^2 p_{3}^2\right)\cr+(m_{3}+m_{4})^2
 \left(-p_{1}^2 p_{2}^2+p_{2}^4-p_{2}^2
   p_{3}^2\right)\Big)\cr
+m_{1}^4 p_{3}^2+m_{1}^2 m_{2}^2 \left(-p_{1}^2+p_{2}^2-p_{3}^2\right)+m_{1}^2 (m_{3}+m_{4})^2
   \left(p_{1}^2-p_{2}^2-p_{3}^2\right)\cr+m_{2}^4 p_{1}^2+m_{2}^2 (m_{3}+m_{4})^2
   \left(-p_{1}^2-p_{2}^2+p_{3}^2\right)+p_{2}^2 (m_{3}+m_{4})^4
 \end{multline}
 and
 \begin{multline}
   c_2(t)=t^2 p_{1}^2 p_{2}^2 p_{3}^2+t\Big(m_{1}^2 \left(-p_{1}^2 p_{3}^2-p_{2}^2 p_{3}^2+p_{3}^4\right)+m_{2}^2 \left(p_{1}^4-p_{1}^2
   p_{2}^2-p_{1}^2 p_{3}^2\right)\cr+(m_{3}-m_{4})^2 \left(-p_{1}^2 p_{2}^2+p_{2}^4-p_{2}^2 p_{3}^2\right)\Big)\cr
    +m_{1}^4 p_{3}^2+m_{1}^2 m_{2}^2 \left(-p_{1}^2+p_{2}^2-p_{3}^2\right)+m_{1}^2 (m_{3}-m_{4})^2
   \left(p_{1}^2-p_{2}^2-p_{3}^2\right)+m_{2}^4 p_{1}^2\cr+m_{2}^2 (m_{3}-m_{4})^2
   \left(-p_{1}^2-p_{2}^2+p_{3}^2\right)+p_{2}^2 (m_{3}-m_{4})^4
 \end{multline}
 and $c_3(t)$ a degree three polynomial which expression is given on the page~\href{https://nbviewer.org/github/pierrevanhove/PicardFuchs/blob/main/PF-icecream-3loop.ipynb}{PF-icecream-3loop}.
 The singularities are located at the roots of
 $(tp_2^2-(m_1+m_2)^2)(tp_2^2-(m_1-m_2)^2) c_1(t) c_2(t)$, the roots
 of $c_3(t)$ are apparent singularities. 

 The homogeneous differential equation
$
\mathcal   L_t^2 f(t)=0   
$
 is a Liouvillian differential equations that satisfies condition of the theorem~11
of~\cite{Fakler}\footnote{We thank Charles Doran for this reference.} and  the solutions are given by
  \begin{equation}
  f^\pm(t)= r(t)^{\frac12}     \exp\left(\pm \int {W(t)\over r(t)} dt\right) 
\end{equation}
with
\begin{align}
  r(t)&={r_2 (p_2^2)^2 t^2+r_1 p_2^2t+ r_0\over (tp_2^2-(m_1+m_2)^2)(tp_2^2-(m_1-m_2)^2)c_1(t)c_2(t)}\\
\noindent        W(t)&=p_2^2{c_3(t)^{5\over2}\over
              q_2(t)^{3\over2}} \left((p_1^2)^2+(p_2^2)^2+(p_3^2)^2-2p_1^2p_2^2-2p_1^2p_3^2-2p_2^2p_3^2\right)
\end{align}
and the coefficients
\begin{align}
              r_2&=(p_1^2)^2+(p_2^2)^2+(p_3^2)^2-2(p_1^2+p_3^2)p_2^2\cr
                   r_1&=\left(-2m_{1}^{2}-2m_{2}^{2}+2m_{3}^{2}+2m_{4}^{2}\right)
                        (p_2^2)^{2}\cr
                        &+\left(\left(4m_{1}^{2}+2m_{2}^{2}-2m_{3}^{2}-2m_{4}^{2}\right)
                          p_1^2 +2 p_3^2
                          \left(m_{1}^{2}+2m_{2}^{2}-m_{3}^{2}-m_{4}^{2}\right)\right)
                          p_2^2 \cr
                          &-2 \left(p_1^2 -p_3^2 \right) \left(m_{1}^{2} p_1^2 -m_{2}^{2} p_3^2 \right)\cr
r_0&=
\left(2m_{3}^{4}+\left(-2m_{1}^{2}-2m_{2}^{2}-4m_{4}^{2}\right)m_{3}^{2}+2m_{4}^{4}+\left(-2m_{1}^{2}-2m_{2}^{2}\right)m_{4}^{2}+m_{1}^{4}+m_{2}^{4}\right)
     (p_2^2)^{2}\cr
     &-2 \left(m_{1} +m_{2} \right) \left(-m_{2} +m_{1} \right)
       \left(m_{1}^{2} p_1^2 -m_{2}^{2} p_3^2
       -\left(m_{3}^{2}+m_{4}^{2}\right) \left(p_1^2 -p_3^2
       \right)\right) p_2^2 \cr
 \nonumber      &+\left(p_1^2 -p_3^2 \right)^{2} \left(-m_{2} +m_{1} \right)^{2} \left(m_{1} +m_{2} \right)^{2}
\end{align}
where $W(t)$ is the Wronskian.
This indicates that we have a period of a rational
surface. This will be proven using a Hodge theoretic analysis for generic
physical parameters in~\cite{DHV}.

 \subsection{The two-scoop ice-cream cone graph}
 \label{sec:2scoop}

\begin{figure}[ht]
  \centering
      \begin{tikzpicture}[baseline=(x)]
        \begin{feynman}
       \vertex (x);
      \vertex[left=2cm of x,label=$p_1$](c1l);
      \vertex[right=2cm of x,label=$p_3$](c1r);
      \vertex[below=3cm of x,label=left:$p_2$,](c1b);
      \tikzfeynmanset{every vertex=dot}
        \vertex [left=1cm  of x] (xl);
        \vertex [right=1cm  of x] (xr);
          \vertex [below=2cm  of x] (xb);
        \diagram* {
         (c1l) -- [fermion] (xl);
         (c1r) -- [fermion] (xr);
         (c1b) -- [fermion] (xb);
         (xl) -- (xr);
     %    (xl) -- (xr);
         (xb) --[out=135,in=-90]  (xl);
          (xb) --[out=45,in=-90] (xr);
				};
                              \end{feynman}
                                           \begin{pgfonlayer}{bg}
                                                  \draw (x) ellipse (1.1cm and .3cm);
                        \end{pgfonlayer}
                   \end{tikzpicture}
            
                   \caption{The two-scoop (three-loop) ice-cream cone graph        \label{fig:icecream2scoop}}
\end{figure}
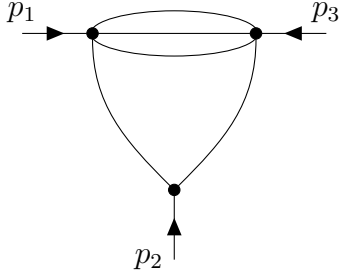
 In this section, we give the Picard--Fuchs operator for 
the two-scoop (three-loop) ice-cream cone graph in
figure~\ref{fig:icecream2scoop} corresponding to the case $n=2$
in~\eqref{e:OmegaIceCream} and~\eqref{e:PolIceCream}.  The results are
summarised on the page~\href{https://nbviewer.org/github/pierrevanhove/PicardFuchs/blob/main/PF-icecream-3loop.ipynb}{PF-icecream-3loop}.

For this case for any
configurations of the internal masses (i.e. with identified masses or
all different masses) we find an irreducible Picard--Fuchs equation  of order
4.  Changing variables from $t$ to $t=1/z$ the differential operator
takes the form  
\begin{equation}
\mathcal  L_z^2=\sum_{r=0}^4 q_r(z) z^r \left(d\over dz\right)^r  \,.
\end{equation}
The operator has for indicial equation $(\rho-2)^2(\rho-3)^2=0$ near $z=0$ (or $t=\infty$) 
which has a two dimension analytic solution near $z=0$ (or $t=\infty$)
with $z^r$ and two non-analytic solutions $z^r\log(z)$ with $r=2,3$.\footnote{This is 
 a Frobenius basis of solutions that can be obtained from the indicial
  equation near $t=0$~\cite{Morrison:1991cd}. The indicial equation
  near the point $t=\alpha$ is 
the equation on the exponents of a solution to the differential
equation behaving as $(t-\alpha)^\rho$. In the following we will
consider $\alpha=0$ or $\alpha=\infty$.    }

It is shown in~\cite{Duhr:2022dxb}, for the all equal masses case, and in~\cite{DHV} for
the generic mass configurations, that  the Picard--Fuchs operator for
the multi-scoop ice-cream cone graphs is fully determined from (two) copies
of the sunset graphs of the scoops. 
 
%%%%%%%%%%%%%%%%%%%%%%%%%%%%%%%%%%%%%%%%%%%%%%%%%%%%%%%%%%%%%%%%%%
\section{Some two-loop graphs differential operator}\label{sec:twoloop}

In this section, we give the Picard--Fuchs operator for some two-loop
graphs in four dimensions with $D=4$ in~\eqref{e:OmeganDef}.
The graph differential form of two-loop graph with $n$ internal edges
in $D$ dimensions is
given by
\begin{equation}
  \Omega_n(t)   = {(\mathcal U_n(\underline x))^{n-{3D\over2}}\over (\mathcal U_n(\underline x)
    \mathcal L_n(\underline m^2,\underline x)-t \mathcal
    V_n(\underline s,\underline x))^{n-D}}\Omega_0^{(n)}
\end{equation}
with $\mathcal U_n(\underline x)$ homogeneous of degree 2, the
kinematic graph polynomial $ \mathcal
    V_n(\underline s,\underline x)$ is homogeneous of degree 3, the mass
hyperplane $\mathcal L_n(\underline x)=\sum_{i=1}^n m_i^2x_i$,
and  $\Omega_0^{(n)}$ the natural differential form on $\mathbb
P^{n-1}$ with coordinates $[x_1:\cdots:x_n]$ as defined
in~\eqref{e:Omega0def}.

The aim of this section is to illustrate the variety of results one
can reach using the algorithm presented earlier.
We consider the following various cases:  the kite graph with $n=5$ in
Section~\ref{sec:kite}, the Tardigrade with $n=6$ in
Section~\ref{sec:Tardigrade}, the double-box graph with $n=7$ in
Section~\ref{sec:DoubleBox} and the pentabox graph with $n=8$ in Section~\ref{sec:Pentabox}.

 \subsection{The kite graph}\label{sec:kite}

\begin{figure}[ht]
  \centering
      \subfloat[]{\label{fig:kite2pt}\begin{tikzpicture}[baseline=(x)]
      \begin{feynman}
       \vertex (x);
      \vertex[above=1cm of x](c1mt);
%      \vertex[below=1cm of x](c1mm);
      \vertex[below=1cm of x](c1mb);
      \vertex[left=1cm of x](c1l);
      \vertex[right=1cm of x](c1r);
   %   \vertex[left=2cm of x,label=$p_1$](xl);
 %     \vertex[right=2cm of x,label=$p_3$](xr);
      \vertex[above=1.5cm of x,label=$p$](xt);
%      \vertex[above=0cm of x,label=$p_2$](xm);
      \vertex[below=1.5cm of x,label=below:$p$](xb);
      \vertex[left=2cm of x](xlphantom);
      \vertex[right=2cm of x](xrphantom);
      \diagram* {
        (c1l) -- (c1mt);
        (c1r) -- (c1mt);
        (c1l) -- (c1r);
        (c1l) -- (c1mb);
        (c1r) -- (c1mb);
    %    (xl) -- [fermion] (c1l);
     %   (xr) -- [fermion] (c1r);
        (xt) -- [fermion] (c1mt);
        (c1mb) -- [fermion] (xb);
  %      (xlphantom) -- [phantom] (xrphantom);
				};
                              \end{feynman}                             
                            \end{tikzpicture}}
                          \qquad
                            \subfloat[]{\label{fig:kite4pt}\begin{tikzpicture}[baseline=(x)]
     \begin{feynman}
       \vertex (x);
      \vertex[above=1cm of x](c1mt);
%      \vertex[below=1cm of x](c1mm);
      \vertex[below=1cm of x](c1mb);
      \vertex[left=1cm of x](c1l);
      \vertex[right=1cm of x](c1r);
      \vertex[left=2cm of x,label=$p_1$](xl);
      \vertex[right=2cm of x,label=$p_3$](xr);
      \vertex[above=1.5cm of x,label=$p_2$](xt);
%      \vertex[above=0cm of x,label=$p_2$](xm);
      \vertex[below=1.5cm of x,label=below:$p_4$](xb);
      \diagram* {
        (c1l) -- (c1mt);
        (c1r) -- (c1mt);
        (c1l) -- (c1r);
        (c1l) -- (c1mb);
        (c1r) -- (c1mb);
        (xl) -- [fermion] (c1l);
        (xr) -- [fermion] (c1r);
        (xt) -- [fermion] (c1mt);
        (xb) -- [fermion] (c1mb);
				};
                              \end{feynman}                   
                      \end{tikzpicture}}
  \caption{kite graph }
  \label{fig:kite}
\end{figure}
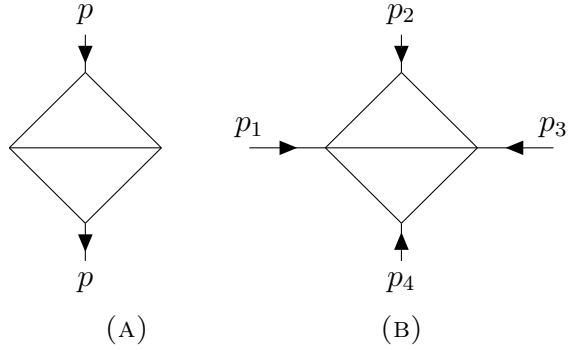

In this section, we consider the kite integral in fig.~\ref{fig:kite}
both in two dimensions and four dimensions. 

 \subsubsection{Triviality of the kite integral in two dimensions}

 In two dimensions,  the rational differential form reads
     \begin{equation}\label{e:Omegakite2d}
  \Omega_{(D=2)\rm kite}^{n-\rm pt}(t) ={ \mathcal U_5(\underline
    x)^{2}\over \mathcal F_{\rm kite}(\underline m^2,\underline
    s^2,t;\underline x)^{3}}\Omega_0^{(5)} 
\end{equation}
with a labelling of the edges where $x_5$ is associated to the middle
line the first Symanzik polynomial reads
 \begin{equation}\label{e:Ukite}
   \mathcal U_5(\underline x)=(x_{1}+x_2)( x_{3}+ x_{4})   +(x_1+x_{2} +x_{3} +x_{4} )x_{5}   
 \end{equation}
and the mass hyperplane
 \begin{equation}\label{e:Lkite}
   \mathcal L_5(\underline m^2,\underline x)=m_1^2x_1+\cdots +m_5^2 x_5   .
 \end{equation}
and the graph polynomial
\begin{equation}\label{e:Fkite}
\mathcal F_{\rm kite}(\underline m^2,\underline
    s^2,t;\underline x)=\mathcal U_5(\underline
    x)\mathcal L_5(\underline
    m^2,\underline x)-t \mathcal V_{\rm kite}^{n-\rm pt}(\underline
    s,\underline x).
  \end{equation}
For the two points graph in fig.~\ref{fig:kite2pt} the kinematic graph  polynomial is given by
 \begin{equation}\label{e:Vkite2pt}
  \mathcal V_{\rm kite}^{2-\rm pt} (\underline s, \underline x) =(x_1+x_4)(x_2+x_3) x_5 +x_1 x_2
  x_3 x_4 \sum_{i=1}^4 {1\over x_i}
\end{equation}
and for the four points graph in fig.~\ref{fig:kite2pt} the kinematic graph  polynomial is given by
 \begin{multline}\label{e:Vkite4pt}
   \mathcal V_{\rm kite}^{4-\rm pt}(\underline s,\underline x)=(p_1+p_2)^2 x_{2} x_{4}
   x_{5}+(p_1+p_4)^2 x_{1} x_{3} x_{5}    \cr
   +p_{1}^2 x_{1} x_{4} x_{5}+p_{2}^2 x_{1} x_{2} (x_{3}+x_{4}+x_{5})+p_{3}^2 x_{2}
   x_{3} x_{5}+p_{4}^2 x_{3} x_{4} (x_{1}+x_{2}+x_{5}),
 \end{multline}
with the momentum conservation condition $p_1+\cdots+p_4=0$ and
$\{p_1,\dots,p_4\}\in\mathbb R^{1,d}$ with $d=1$ (in two dimensions)
or $d=3$ (in four dimensions).

We first explain that in two dimensions the differential for the kite
integral is trivial, and give its consequences. A proof of the
triviality of the differential form is given afterwards.

In two-dimensions both the two- and four-points cases the integral is trivial
because $\mathcal U_5(\underline x)^{2}$ lies in the Jacobian ideal of
$\mathcal F_{\rm kite}(\underline m^2,\underline
s,t;\underline x)$ because
\begin{equation}\label{e:Ured}
    \mathcal U_5(\underline x)^{2}=\sum_{i=1}^5C_i(\underline
      x)  \partial_{x_i} \mathcal F_{\rm kite}(\underline m^2,\underline
s^2,t;\underline x).
\end{equation}
There is a choice of coefficients  $C_i(\underline x)$ such that
\begin{equation}\label{e:dC}
  \sum_{i=1}^5\partial_{x_i} C_i(\underline x) =0.
\end{equation}
This implies that
\begin{equation}
    { \mathcal U_5(\underline x)^{2}\over \mathcal F_{\rm kite}(\underline m^2,\underline
s^2,t;\underline x)^3}= -\frac12 \sum_{i=1}^5 \partial_{x_i} {C_i(\underline x) \over \mathcal F_{\rm kite}(\underline m^2,\underline
s^2,t;\underline x)^2}.
\end{equation}
This implies that the differential form in~\eqref{e:Omegakite2d} is
trivial
\begin{equation}
  \Omega_{(D=2)\rm kite}^{n-\rm pt}(t) =d\beta_{(D=2)\rm kite}^{n-\rm pt}\,.
\end{equation}
As a consequence, the kite integral Feynman in $D=2$ reduces to its
boundary components given by a sum of four one-scoop ice-cream cone like Feynman
integrals (from the boundaries at $x_i=0$ with $1\leq i\leq 4$) and
a product of two one-loop integral from $x_5=0$. There is no
contributions from $x_i=\infty$ with $1\leq i\leq 5$ as can be checked
on the expressions given on the 
page~\href{https://nbviewer.org/github/pierrevanhove/PicardFuchs/blob/main/PF-Kite.ipynb}{PF-Kite}.

We give a proof of the reduction in~\eqref{e:Ured} with the
condition~\eqref{e:dC}.  The coefficients $C_i(\underline x)$ are
homogeneous polynomial of degree 2 in the variables $\underline
x=\{x_1,\dots,x_5\}$
\begin{equation}
  C_i(\underline x)= \sum_{a_1+\cdots+a_5=0\atop 0\leq a_i\leq 2} c_{i}(a_1,\dots,a_5) \prod_{r=1}^5 x_r^{a_r}  .
\end{equation}
Each polynomial has 15 coefficients.  The condition~\eqref{e:dC} leads
to five linear equations, that we can solve for, say $c_5(0, 0, 0, 0, 2)$, $c_5(0, 0, 0, 1, 1)$, $c_5(0, 0, 1, 0, 1)$, 
$c_5(0, 1, 0, 0, 1)$, $c_5(1, 0, 0, 0, 1)$.
We set to zero the coefficients
\begin{multline}
 c_2(1, 0, 1, 0, 0)=c_3(0, 1, 1, 0, 0)=c_3(0, 2, 0, 0, 
 0)
 =c_3(1, 0, 0, 1, 0) \cr
 =c_3(1, 0, 1, 0, 0)=c_3(1, 1, 0, 
 0, 0)=c_3(2, 0, 0, 0, 0)
 =c_4(0, 1, 1, 0, 0)\cr
 =c_4(0, 2, 
 0, 0, 0)=c_4(1, 0, 0, 1, 0)=c_4(1, 0, 1, 0, 0)
 =c_4(1, 
   1, 0, 0, 0) \cr=c_4(2, 0, 0, 0, 0)=c_5(0, 1, 1, 0, 0)=c_5(
   0, 2, 0, 0, 0)
   =c_5(1, 0, 0, 1, 0) \cr=c_5(1, 0, 1, 0, 
   0)=c_5(1, 1, 0, 0, 0)=c_5(2, 0, 0, 0, 0)=0  .
 \end{multline}
The Jacobian reduction   in~\eqref{e:Ured} leads to 64 linear
equations in the remaining 51 coefficients of the polynomial $C_i$,
denoted $\vec C$,
\begin{equation}
A  \vec C   = \vec B
\end{equation}
with $B^T=\{0, 0, 1, 0, 0, 0, 2, 0, 0, 1, 0, 0, 0, 0, 0, 0, 2, 2, 0, 2, 4, 0, 2, 
0, 0, 1, 2,$ $ 1, 2, 2, 1, 0, 0, 0, 0, 2, 2, 0, 2, 4, 0, 2, 0, 0, 2, 4, 
2, 4, 4, 2, 0, 0, 0, 1, 2, 1, 2, 2, 1, 0,$  $0, 0, 0, 0\}$.
The matrix $A$ has rank 47 for the two-point case, 
and rank  51 for the
four-point case. 
The system has a unique solution in the two-point case, but it  has a unique solution only when the momenta
$\{p_1,\dots,p_4\}\in\mathbb R^{1,1}$ are in two dimensions. If the
momenta are taken in four dimensions $\{p_1,\dots,p_4\}\in\mathbb
R^{1,3}$,  there
is no solution to the system, and the integral is not any more trivial.

\subsubsection{The kite integral in four dimensions}\label{sec:Kite4D}

In this section, we consider the kite integral in four dimensions
$D=4$. The rational  differential form in $\mathbb P^4$ associated to the
massive two- and four-point kite graph, in fig.~\ref{fig:kite}, in
four-dimensions read
\begin{equation}\label{e:Omegakite}
  \Omega_{\rm kite}^{n-\rm pt}(t) ={ \Omega_0^{(5)} \over \mathcal U_5(\underline x)
    (\mathcal U_5(\underline
    x) \mathcal L_5(\underline
    m^2,\underline x)-t \mathcal V_{\rm kite}^{n-\rm pt}(\underline
    s,\underline x))}
 \end{equation}
 with the $\mathcal U_5(\underline x)$ given in~\eqref{e:Ukite} and
 the mass hyperplane $\mathcal L_5(\underline m^2,\underline x)$ given in~\eqref{e:Lkite}.

\noindent $\bullet$  We start considering  the case of the two-point massive kite graph of
figure~\ref{fig:kite2pt} studied in~\cite{Broadhurst:1987ei} by
 dispersion relations in $D=4$ dimensions.  In this case the kinematic
 graph polynomial is given in~\eqref{e:Vkite2pt}.
In $D=4$ the  Picard--Fuchs operator is given by 
 \begin{equation}
   \mathcal L_{ (D=4)\rm kite}^{2-\rm pt}= t{d\over dt}+1   \,.
 \end{equation}
 The Feynman integral satisfies the inhomogeneous differential
 equation 
 \begin{equation}
      \mathcal L_{\rm kite}^{2-\rm pt}\int_{x_i\geq0} \Omega_{\rm
        kite}^{2-\rm pt}(t) = \sum_{i=1}^4 S_i(t).
    \end{equation}
where $S_i(t)$ are two-loop sunset like contribution of
Section~\ref{sec:2sunset}, see the
page~\href{https://nbviewer.org/github/pierrevanhove/PicardFuchs/blob/main/PF-Kite.ipynb}{PF-Kite} for the expression for
generic mass configurations.
 This result generalises the special cases with vanishing internal
 masses
 considered
 in~\cite{Remiddi:2016gno,Adams:2016xah,Bogner:2017vim,Bogner:2018uus,
   Broedel:2018qkq,Bezuglov:2020ywm} derived using other methods.

\noindent $\bullet$ We now turn to the four-point case of fig.~\ref{fig:kite4pt} which
has for kinematic graph polynomial is given in~\eqref{e:Vkite4pt}.
 On the numerical cases studied
 on~\href{https://nbviewer.org/github/pierrevanhove/PicardFuchs/blob/main/PF-Kite.ipynb}{PF-Kite},  the algorithm gives seven order operators that 
 factorise using the factorisation algorithm~\cite{facto} into the
 product of four differential operators
 \begin{equation}
   \mathcal L_{\rm kite}^{4-\rm pt}= L^{(1)}L^{(2)}L^{(3)}
   \left(z{d\over dz}+1\right)
 \end{equation}
 where $L^{(r)}$ with $r=1,2,3$ are order~2 operators.
 Although the factorisation is not unique, we still use the
 one obtained to determine the type of solutions of the differential operator.
 All the factors have only Liouvillian solutions, as computed by Maple.
 This can also be checked using a criterion by Falker in  Theorem~11
of~\cite{Fakler}.
 Therefore, the product operator $\mathcal L_{\rm kite}^{4-\rm pt}$ also have only Liouvillian solutions.\footnote{
   The solutions of a product operator~$AB$ are solutions of the inhomogeneous equation~$B(y) = u$, where~$u$ is a solution of~$A(u) = 0$. Using variation of parameters, the solutions of this inhomogeneous equations can be expressed in terms of the solutions of~$B$ and that of~$A$, using only the operations defining Liouvillian functions.}
This is explained using a Hodge theoretic analysis for generic
physical parameters in~\cite[Remark~6.11]{DHV}.
 
   \subsection{The tardigrade}\label{sec:Tardigrade}

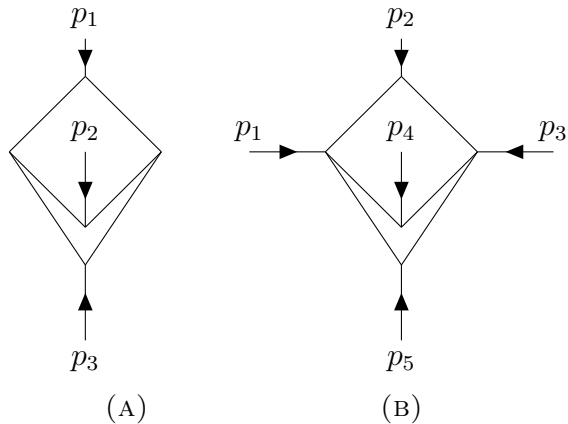
\begin{figure}[ht]
  \centering
      \subfloat[]{\label{fig:tardi3pt}\begin{tikzpicture}[baseline=(x)]
      \begin{feynman}
       \vertex (x);
      \vertex[above=1cm of x](c1mt);
      \vertex[below=1cm of x](c1mm);
      \vertex[below=1.5cm of x](c1mb);
      \vertex[left=1cm of x](c1l);
      \vertex[right=1cm of x](c1r);
   %   \vertex[left=2cm of x,label=$p_1$](xl);
 %     \vertex[right=2cm of x,label=$p_3$](xr);
      \vertex[above=1.5cm of x,label=$p_1$](xt);
      \vertex[above=0cm of x,label=$p_2$](xm);
      \vertex[below=2.5cm of x,label=below:$p_3$](xb);
      \diagram* {
        (c1l) -- (c1mt);
        (c1r) -- (c1mt);
        (c1l) -- (c1mm);
        (c1r) -- (c1mm);
        (c1l) -- (c1mb);
        (c1r) -- (c1mb);
    %    (xl) -- [fermion] (c1l);
     %   (xr) -- [fermion] (c1r);
        (xt) -- [fermion] (c1mt);
        (xm) -- [fermion] (c1mm);
        (xb) -- [fermion] (c1mb);
				};
                              \end{feynman}                             
                            \end{tikzpicture}}
                          \qquad
                            \subfloat[]{\label{fig:tardi5pt}\begin{tikzpicture}[baseline=(x)]
        \begin{feynman}
       \vertex (x);
      \vertex[above=1cm of x](c1mt);
      \vertex[below=1cm of x](c1mm);
      \vertex[below=1.5cm of x](c1mb);
      \vertex[left=1cm of x](c1l);
      \vertex[right=1cm of x](c1r);
      \vertex[left=2cm of x,label=$p_1$](xl);
      \vertex[right=2cm of x,label=$p_3$](xr);
      \vertex[above=1.5cm of x,label=$p_2$](xt);
      \vertex[above=0cm of x,label=$p_4$](xm);
      \vertex[below=2.5cm of x,label=below:$p_5$](xb);
      \diagram* {
        (c1l) -- (c1mt);
        (c1r) -- (c1mt);
        (c1l) -- (c1mm);
        (c1r) -- (c1mm);
        (c1l) -- (c1mb);
        (c1r) -- (c1mb);
        (xl) -- [fermion] (c1l);
        (xr) -- [fermion] (c1r);
        (xt) -- [fermion] (c1mt);
        (xm) -- [fermion] (c1mm);
        (xb) -- [fermion] (c1mb);
				};
                              \end{feynman}                              
                      \end{tikzpicture}}
  \caption{Tardigrade graph }
  \label{fig:tardigrade}
\end{figure}

The tardigrade graphs in fig~\ref{fig:tardigrade} have the rational
differential form in $\mathbb P^5$

\begin{equation}
  \Omega_{\rm Tardigrade}^{n-\rm pt}(t) ={\Omega_0^{(6)} \over
   \left (\mathcal U_6(\underline x)
    \mathcal L_6(\underline m^2,\underline x)-t \mathcal V_{\rm
      Tardigrade}^{n-\rm pt}(\underline s,\underline x)\right)^2}
 \end{equation}
 with
 \begin{equation}
   \mathcal U_6(\underline x)=(x_1 + x_2) (x_3 + x_4) + (x_1 + x_2) (x_5 + x_6) + (x_3 + x_4) (x_5 + x_6)
 \end{equation}
 the mass hyperplane
 \begin{equation}
   \mathcal L_6(\underline m^2,\underline x)=m_1^2x_1+\cdots +m_6^2 x_6.
 \end{equation}

\subsubsection{The three points case} For the  three points tardigrade in
 fig.~\ref{fig:tardi3pt} the kinematic graph polynomial is given by 
 \begin{multline}
   \mathcal V_{\rm Tardigrade}^{3-\rm pt}(\underline s,\underline x)=p_{1}^2 (x_{1} x_{2} (x_{3}+x_{4}+x_{5}+x_{6})+x_{1} x_{4} x_{6}+x_{2} x_{3}
   x_{5})\cr +p_{2}^2 x_{4} (x_{1} (x_{3}+x_{5})+x_{2} (x_{3}-x_{6})+x_{3}
   (x_{5}+x_{6}))\cr
   +p_{3}^2 (x_{1} x_{6} (x_{3}+x_{5})+x_{2} x_{6} (x_{3}+x_{5})+x_{2}
   x_{4} (x_{5}+x_{6})+x_{5} x_{6} (x_{3}+x_{4}))
 \end{multline}
 with $\{p_1,\dots,p_3\}\in\mathbb R^{1,3}$ and $p_1+p_2+p_3=0$.
 For generic values of the mass parameters and external momenta, the
 algorithm gives a reducible order 3  Picard--Fuchs  operator which is
 minimal order Picard--Fuchs  operator $\mathcal L_t$
 is of order 2 in $t$. Near $t=0$ with one analytic solution
 behaving as $1/t$ and one logarithmic solution behaving as
 $\log(t)/t$. Various  numerical cases with the analytic solution are given on the page~\href{https://nbviewer.org/github/pierrevanhove/PicardFuchs/blob/main/PF-Tardigrade.ipynb}{PF-Tardigrade}.

In the space case of all equal masses
 $m_1=\cdots=m_6$ and all equal kinematics 
 $p_1^2=p_2^2=p_3^2$. This  is a single scale problem, which can be
 reabsorbed  by redefining $t \to t m_1^2/p_1^2$ such that the
 rational differential form becomes
 \begin{equation}
  \Omega_{\rm Tardigrade}^{3-\rm pt}(t) ={\Omega_0^{(6)} \over m_1^2
    \left(\mathcal U_6(\underline x)
    \sum_{i=1}^6x_i -t \mathcal V_{\rm
      Tardigrade}^{3-\rm pt}(\underline x)\right)^2}
 \end{equation}

 Then kinematic  graph polynomial reads
 \begin{multline}
        \mathcal V_{\rm Tardigrade}^{3-\rm pt}(\underline
        x)=x_{1} (x_{2} x_{3}+x_{3} (x_{4}+x_{5}+x_{6})+x_{4}
   x_{6})+x_{1} x_{4} (x_{2}+x_{5})+x_{1} x_{6}
   (x_{2}+x_{5})\cr+x_{2} x_{3} (x_{4}+x_{6}+x_5)
   +x_{5} x_{6} (x_{2}+x_{4})+x_{2} x_{4}
   (x_{5}+x_{6})+x_{3} x_{5} (x_{4}+x_{6}).
 \end{multline}
In this   the
minimal order Picard--Fuchs operator is of order 1
\begin{equation}
  \mathcal L_t =t(t - 4)
  {d\over dt} + 2t - 6.
\end{equation}

\subsubsection{The five points case}
 For the five points case  in
fig.~\ref{fig:tardi5pt}  the kinematic graph polynomial is given by 
 \begin{multline}
   \mathcal V_{\rm Tardigrade}^{5-\rm pt}(\underline s,\underline
        x)=p_{1}^2 x_{1} x_{5} (x_{3}+x_{4})+p_{2}^2 x_{1} x_{2}
   (x_{3}+x_{4}+x_{5}+x_{6})+p_{3}^2 x_{2} x_{4} (x_{5}+x_{6})\cr
   +p_{4}^2 x_{4} (x_{1}
   (x_{3}+x_{5})+x_{2} (x_{3}+x_{5})+x_{3} (x_{5}+x_{6}))\cr
   +p_{5}^2 x_{5} (x_{1}
   (x_{4}+x_{6})+x_{2} (x_{4}+x_{6})+x_{6} (x_{3}+x_{4}))\cr
   +(p_1+p_2)^2 x_{2} x_{5}
   (x_{3}+x_{4})+(p_2+p_3)^2 x_{1} x_{4} (x_{5}+x_{6})+(p_3+p_4)^2 x_{2} (x_{3} x_{6}-x_{4}
   x_{5})\cr
   -(p_4+p_5)^2 x_{4} x_{5} (x_{1}+x_{2})+(p_5+p_1)^2 x_{1} (x_{3} x_{6}-x_{4} x_{5})  
 \end{multline}
with $\{p_1,\dots,p_5\}\in\mathbb R^{1,3}$ and $p_1+\cdots+p_5=0$.
Depending on the    configuration of the external momenta, the order of the
differential operator is between 6 and 11.

\begin{itemize}
\item For instance the symmetric case with all equal masses
$m_1^2=\cdots=m_6^2$ and
$p_1^2=p_2^2=p_3^2=p_4^2=(p_1+p_2)^2=(p_1+p_4)^2$, by rescaling the
$t$ parameter  by $t\to t m_1^2/p_1^2$, the kinematic graph polynomial becomes
\begin{multline}
  V_{\rm Tardigrade}^{5-\rm pt}(\underline s,\underline
        x)=x_{1} (x_{2} (x_{3}+x_{4}+x_{5}+x_{6})+x_{3}
  (x_{4}+x_{5}+x_{6})+x_{5} (x_{4}+x_{6}))\cr
  +x_{2} x_{3}
   (x_{4}+x_{5})+x_{2} x_{6} (x_{5}-2 x_{4})+x_{3} x_{4}
   x_{5}+x_{3} x_{4} x_{6}+x_{3} x_{5} x_{6}+x_{4}
   x_{5} x_{6}  
\end{multline}
the algorithm gives a Picard--Fuchs operator is of order 6 that is not
factorised by the factorisation algorithm~\cite{facto}  with an head
polynomial of degree 50 in $t$.

\item When the masses $m_i$ with $1\leq i\leq 6$  parameters and the
kinematics parameters $p_1^2$ , $p_2^2$,  $p_3^2$, $p_4^2$,
$(p_1+p_2)^2$, $(p_1+p_4)^2$ take generic values the algorithm gives a
Picard--Fuchs operator is of order 11 with an head polynomial of
degree up to 215.
We have checked using the factorisation
  algorithm~\cite{facto} that these differential operators
  do not factorise. These results are
 compatible with the conjecture that the tardigrade Feynman
 integrals are $K3$ periods
 integrals~\cite{Bourjaily:2018yfy,Bourjaily:2019hmc}.  For the case
 of vanishing masses $m_1=\cdots=m_6=0$, the integral will develop
 new singularities and the order of the Picard--Fuchs operator is expected to decrease.
\end{itemize}

\subsection{The double-box graphs}\label{sec:DoubleBox}

\begin{figure}[ht]
  \centering
      \subfloat[]{\label{fig:dbox4pt}\begin{tikzpicture}[baseline=(x)]
        \begin{feynman}
       \vertex (x);
      \vertex[above=1cm of x](c1mt);
      \vertex[below=1cm of x](c1mb);
      \vertex[above left=2.5cm of x,label=$p_2$](x1lt);
      \vertex[below left=2.5cm of x,label=$p_1$](x1lb);
      \vertex[above right=2.5cm of x,label=$p_3$](x1rt);
      \vertex[below right=2.5cm of x,label=$p_4$](x1rb);
%      \tikzfeynmanset{every vertex=dot}
      \vertex[above left=1.4cm of x](c1lt);
       \vertex[below left=1.4cm of x](c1lb);
       \vertex[above right=1.4cm of x](c1rt);
       \vertex[below right=1.4cm of x](c1rb);
        \diagram* {
         (c1lt) -- (c1rt);
         (c1lb) -- (c1rb);
         (c1lt) -- (c1lb);
         (c1mt) --  (c1mb);
         (c1rt) --(c1rb);
         (x1lt) -- [fermion] (c1lt);
         (x1lb) -- [fermion] (c1lb);
         (x1rt) -- [fermion] (c1rt);
         (x1rb) -- [fermion] (c1rb);

				};
                              \end{feynman}                              
                            \end{tikzpicture}}
                          \qquad
                            \subfloat[]{\label{fig:dbox6pt}\begin{tikzpicture}[baseline=(x)]
        \begin{feynman}
       \vertex (x);
      \vertex[above=1cm of x](c1mt);
      \vertex[below=1cm of x](c1mb);
       \vertex[above=2cm of x,label=$p_3$](x1mt);
      \vertex[below=2cm of x,label=below:$p_6$](x1mb);
      \vertex[above left=2.5cm of x,label=$p_2$](x1lt);
      \vertex[below left=2.5cm of x,label=$p_1$](x1lb);
      \vertex[above right=2.5cm of x,label=$p_4$](x1rt);
      \vertex[below right=2.5cm of x,label=$p_5$](x1rb);
%      \tikzfeynmanset{every vertex=dot}
      \vertex[above left=1.4cm of x](c1lt);
       \vertex[below left=1.4cm of x](c1lb);
       \vertex[above right=1.4cm of x](c1rt);
       \vertex[below right=1.4cm of x](c1rb);
        \diagram* {
         (c1lt) -- (c1rt);
         (c1lb) -- (c1rb);
         (c1lt) -- (c1lb);
         (c1mt) --  (c1mb);
         (c1rt) --(c1rb);
         (x1lt) -- [fermion] (c1lt);
         (x1lb) -- [fermion] (c1lb);
         (x1rt) -- [fermion] (c1rt);
         (x1rb) -- [fermion] (c1rb);
         (x1mt) -- [fermion] (c1mt);
          (x1mb) -- [fermion] (c1mb);
				};
                              \end{feynman}                              
                      \end{tikzpicture}}
  \caption{Double box graphs.}
  \label{fig:doublebox}
\end{figure}
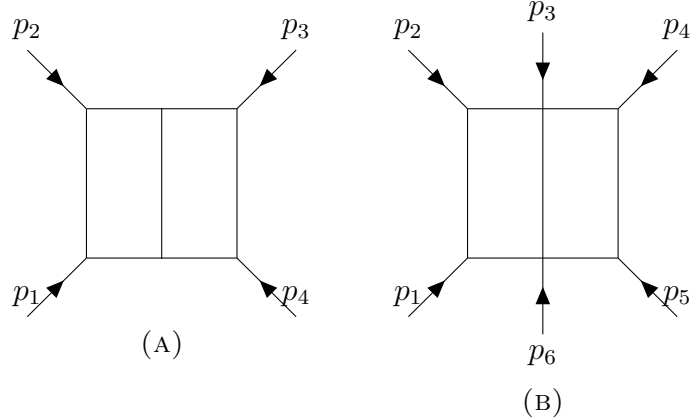

For the double box graph in fig.~\ref{fig:doublebox} we have a
rational differential form in $\mathbb P^6$
\begin{equation}
  \Omega_{\rm Double-box}^{n-\rm pt}(t) ={\mathcal U_7(\underline x)  \over
    (\mathcal U_7(\underline x)\mathcal L_7(\underline m^2,\underline
    x)-t \mathcal V_{\rm DoubleBox}^{n-\rm pt}(\underline s,\underline
    x))^3}\Omega_0^{(7)}
 \end{equation}
 with a labelling of the edges where the middle internal edge is
 labelled by the variables $x_7$ the first Symanzik polynomial reads
 \begin{equation}
   \mathcal U_7(\underline x)=(x_1+x_2+x_3)(x_4+x_5+x_6)+(x_1+\cdots+x_6)x_7
 \end{equation}
 the mass hyperplane
 \begin{equation}
   \mathcal L_7(\underline m^2,\underline x)=m_1^2x_1+\cdots +m_7^2 x_7
 \end{equation}
\noindent $\bullet$ 
 For the four points graph in fig.~\ref{fig:dbox4pt} the kinematic
 graph polynomial reads
 \begin{multline}
   \mathcal V_{\rm DoubleBox}^{4-\rm pt}(\underline s,\underline x)=p_{1}^2 x_{2} (x_{1} (x_{4}+x_{5}+x_{6}+x_{7})+x_{6} x_{7})+p_{2}^2 x_{2} (x_{3}
   (x_{4}+x_{5}+x_{6}+x_{7})+x_{4} x_{7})\cr
   +p_{3}^2 x_{5} (x_{1} x_{4}+x_{2}
   x_{4}+x_{3} x_{4}+x_{3} x_{7}+x_{4} x_{7})+p_{4}^2 x_{5} (x_{1}
   (x_{6}+x_{7})+x_{6} (x_{2}+x_{3}+x_{7}))\cr
   +(p_1+p_2)^2 (x_{1} x_{3}
   (x_{4}+x_{5}+x_{6}+x_{7})+x_{1} x_{4} (x_{6}+x_{7})+x_{6} (x_{2} x_{4}+x_{3}
   (x_{4}+x_{7})+x_{4} x_{7}))\cr
   +(p_1+p_4)^2 x_{2} x_{5} x_{7}
 \end{multline}
 with $\{p_1,\dots,p_4\}\in\mathbb R^{1,3}$ and $p_1+\cdots +p_4=0$.

 \noindent $\bullet$
 For the six points graph in fig.~\ref{fig:dbox6pt}  the kinematic
 graph polynomial reads
 \begin{multline}
   \mathcal V_{\rm DoubleBox}^{6-\rm pt}(\underline s,\underline x)=p_{1}^2 x_{1} x_{2} (x_{4}+x_{5}+x_{6}+x_{7})+p_{2}^2
   x_{2} x_{3} (x_{4}+x_{5}+x_{6}+x_{7})+p_{3}^2 x_{3}
   x_{6} x_{7}\cr
   +p_{4}^2 x_{5} x_{6}
   (x_{1}+x_{2}+x_{3}+x_{7})+p_{5}^2 x_{4} x_{5}
   (x_{1}+x_{2}+x_{3}+x_{7})+p_{6}^2 x_{1} x_{4}
   x_{7}\cr
   +(p_1+p_2)^2 x_{1} x_{3}
   (x_{4}+x_{5}+x_{6}+x_{7})
   +(p_1+p_2+p_3)^2 x_{1} x_{6}
   x_{7}+(p_2+p_3)^2 x_{2} x_{6} x_{7}
   \cr+(p_2+p_3+p_4)^2 x_{2} x_{5}
   x_{7}
   +(p_3+p_4)^2 x_{3} x_{5} x_{7}+(p_3+p_4+p_5)^2 x_{3} x_{4}
   x_{7}\cr+(p_4+p_5)^2 x_{4} x_{6}
   (x_{1}+x_{2}+x_{3}+x_{7})+(p_5+p_6)^2 x_{1} x_{5}
   x_{7}+(p_6+p_1)^2 x_{2} x_{4} x_{7}
 \end{multline}
 with $\{p_1,\dots,p_6\}\in\mathbb R^{1,3}$ and $p_1+\cdots +p_6=0$.
In the six points case we impose the Gram determinant conditions
 listed in~\cite{Asribekov:1962tgp} by
 taking all the kinematics in four dimensions.   

We find that both the four- and six-point  massive double-box integrals (see
figure~\ref{fig:dbox4pt} and~\ref{fig:dbox6pt}) have a Fuchsian  differential operator of order
2 with only regular singularities. We find 
Picard-Fuchs operators of the form
\begin{equation}
  \mathcal L_t^{n-\rm  pt}= q_2(t) \left( t {d\over dt}\right)^2+
  q_1(t) t {d\over dt}+ q_0(t) \,,
\end{equation}
where $q_2(t)$ has single roots different from 0.
At $t=0$ the indicial equation is $(\rho+1)^2=0$ for the six point case, therefore a local  basis of solutions behaves as $1/t$ and
$\log(t)/t$. Showing
that the space of analytic solutions near $t=0$ is one dimensional, and there is
a logarithmic contribution. This is compatible with the fact that the maximal cut leads to period
  integral for an 
  elliptic
  curve~\cite{Caron-Huot:2012awx,Bloch:2021hzs,Bourjaily:2017bsb,Pozo:2022dox}.

If one does not impose the kinematic conditions implied by the Gram
determinant conditions (which means relaxing the condition that 
the external momenta in a  four dimensional space), the Picard--Fuchs
operator of order 4. It is not factorisable as certified by the factorisation
algorithm~\cite{facto}. At $t=0$  the indicial equation is
$\rho(\rho-1)(\rho+1)^2=0$ therefore a local basis of solutions behaves
as  $1$, $t$ and $1/t$ and $\log(t)/t$. Showing
that the space of analytic solutions near $t=0$ is three-dimensional, and there is
a logarithmic contribution. 
This illustrates  how the kinematic
relations impose relations between the coefficients of the graph
polynomial and affect the singularity structure of the rational
differential form. We see that the Gram condition reduces the number
of analytic solution near $t=0$.
This will be proven using a Hodge theoretic analysis for generic
physical parameters in~\cite{DHV}.

These results are given on the page~\href{https://nbviewer.org/github/pierrevanhove/PicardFuchs/blob/main/PF-DoubleBox.ipynb}{PF-DoubleBox}.

\subsection{The pentabox graphs}\label{sec:Pentabox}

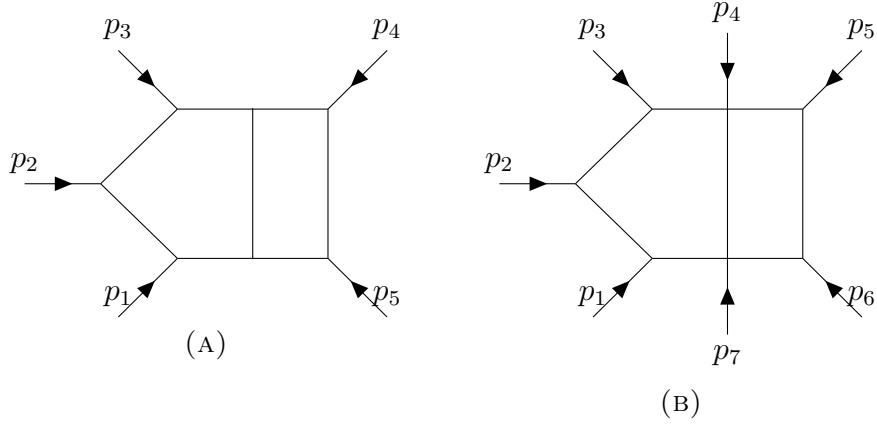
\begin{figure}[ht]
  \centering
      \subfloat[]{\label{fig:pentabox5pt}\begin{tikzpicture}[baseline=(x)]
        \begin{feynman}
       \vertex (x);
      \vertex[above=1cm of x](c1mt);
      \vertex[below=1cm of x](c1mb);
     \vertex[below left=2.5cm of x,label=$p_1$](x1lb);
     \vertex[left=3cm of x,label=$p_2$](x1lm);
     \vertex[above left=2.5cm of x,label=$p_3$](x1lt);
      \vertex[above right=2.5cm of x,label=$p_4$](x1rt);
      \vertex[below right=2.5cm of x,label=$p_5$](x1rb);
%      \tikzfeynmanset{every vertex=dot}
      \vertex[above left=1.4cm of x](c1lt);
       \vertex[left=2cm of x](c1lm);
       \vertex[below left=1.4cm of x](c1lb);
       \vertex[above right=1.4cm of x](c1rt);
       \vertex[below right=1.4cm of x](c1rb);
        \diagram* {
         (c1lt) -- (c1rt);
         (c1lb) -- (c1rb);
         (c1lt) -- (c1lm);
         (c1lb) -- (c1lm);
         (c1mt) --  (c1mb);
         (c1rt) --(c1rb);
         (x1lt) -- [fermion] (c1lt);
         (x1lm) -- [fermion] (c1lm);
         (x1lb) -- [fermion] (c1lb);
         (x1rt) -- [fermion] (c1rt);
         (x1rb) -- [fermion] (c1rb);

				};
                              \end{feynman}                              
                            \end{tikzpicture}}
                          \qquad
                            \subfloat[]{\label{fig:pentabox7pt}\begin{tikzpicture}[baseline=(x)]
        \begin{feynman}
       \vertex (x);
       \vertex[above=1cm of x](c1mt);
      \vertex[below=1cm of x](c1mb);
      \vertex[above=2cm of x,label=$p_4$](x1mt);
      \vertex[left=3cm of x,label=$p_2$](x1lm);
      \vertex[below=2cm of x,label=below:$p_7$](x1mb);
      \vertex[above left=2.5cm of x,label=$p_3$](x1lt);
      \vertex[below left=2.5cm of x,label=$p_1$](x1lb);
      \vertex[above right=2.5cm of x,label=$p_5$](x1rt);
      \vertex[below right=2.5cm of x,label=$p_6$](x1rb);
%      \tikzfeynmanset{every vertex=dot}
      \vertex[above left=1.4cm of x](c1lt);
       \vertex[left=2cm of x](c1lm);
       \vertex[below left=1.4cm of x](c1lb);
       \vertex[above right=1.4cm of x](c1rt);
       \vertex[below right=1.4cm of x](c1rb);
        \diagram* {
          (c1lt) -- (c1rt);
         (c1lb) -- (c1rb);
         (c1lt) -- (c1lm);
         (c1lb) -- (c1lm);
         (c1mt) --  (c1mb);
         (c1rt) --(c1rb);
         (x1lt) -- [fermion] (c1lt);
          (x1lm) -- [fermion] (c1lm);
         (x1lb) -- [fermion] (c1lb);
         (x1rt) -- [fermion] (c1rt);
         (x1rb) -- [fermion] (c1rb);
         (x1mt) -- [fermion] (c1mt);
          (x1mb) -- [fermion] (c1mb);
				};
                              \end{feynman}                              
                      \end{tikzpicture}}
  \caption{The Pentabox graphs.}
  \label{fig:pentabox}
\end{figure}
For the Pentabox graph in fig.~\ref{fig:pentabox} we have a
rational differential form in $\mathbb P^7$

\begin{equation}
  \Omega_{\rm Pentabox}^{n-\rm pt}(t) ={\mathcal U_8(\underline x) ^2 \over
    \left(\mathcal U_8(\underline x)\mathcal L_8(\underline m^2,\underline
    x)-t \mathcal V_{\rm PentaBox}^{n-\rm pt}(\underline s,\underline x)\right)^4}\Omega_0^{(8)}
 \end{equation}
 with a labelling of the edges where the middle internal edge is
 labelled by the variables $x_8$ the first Symanzik polynomial reads
 \begin{equation}
   \mathcal U_8(\underline x)=(x_1+x_2+x_3+x_4)(x_5+x_6+x_7)+x_8(x_1+\cdots+x_7)
 \end{equation}
 the mass hyperplane
 \begin{equation}
   \mathcal L_8(\underline m^2,\underline x)=m_1^2x_1+\cdots +m_8^2 x_8.
 \end{equation}

 \noindent $\bullet$ For the five-point case of figure~\ref{fig:pentabox5pt} the
 kinematic graph
 polynomial is given by
 \begin{multline}
   \mathcal V_{\rm Pentabox}^{5-\rm pt}(\underline s,\underline x)=p_{1}^2 x_{2} (x_{1} (x_{5}+x_{6}+x_{7}+x_{8})+x_{5}
   x_{8})+p_{2}^2 x_{2} x_{3}
   (x_{5}+x_{6}+x_{7}+x_{8})\cr
   +p_{3}^2 x_{3} (x_{4}
   (x_{5}+x_{6}+x_{7}+x_{8})+x_{7} x_{8})+p_{4}^2 x_{6}
   (x_{1} x_{7}+x_{2} x_{7}+x_{3} x_{7}+x_{4}
   x_{7}+x_{4} x_{8}+x_{7} x_{8})\cr
   +p_{5}^2 x_{6} (x_{1}
   (x_{5}+x_{8})+x_{5} (x_{2}+x_{3}+x_{4}+x_{8}))+(p_1+p_2)^2
   x_{3} (x_{1} (x_{5}+x_{6}+x_{7}+x_{8})+x_{5}
   x_{8})\cr
   +(p_2+p_3)^2 x_{2} (x_{4}
   (x_{5}+x_{6}+x_{7}+x_{8})+x_{7} x_{8})+(p_3+p_4)^2 x_{3}
   x_{6} x_{8}\cr
   +(p_4+p_5)^2 (x_{1} x_{4}
   (x_{5}+x_{6}+x_{7}+x_{8})+x_{1} x_{7}
   (x_{5}+x_{8})+x_{5} (x_{2} x_{7}+x_{3} x_{7}+x_{4}
   x_{7}+x_{4} x_{8}+x_{7} x_{8}))\cr
   +(p_5+p_1)^2 x_{2} x_{6}
   x_{8} ,
 \end{multline}
 with $\{p_1,\dots,p_5\}\in\mathbb R^{1,3}$ and $p_1+\cdots +p_5=0$.

\noindent $\bullet$ For the seven-point case of fig.~\ref{fig:pentabox7pt} the kinematic graph
 polynomial is given by 
 \begin{multline}
   \mathcal V_{\rm PentaBox}^{7-\rm pt}(\underline s,\underline x)=p_{1}^2 x_{2} (x_{1} x_{5}+x_{1} x_{6}+x_{1}
   x_{7}+x_{1} x_{8}+x_{5} x_{8}+x_{6} x_{8}+x_{7}
   x_{8}) \cr
   +p_{2}^2 x_{2} x_{3}
   (x_{5}+x_{6}+x_{7}+x_{8})
   +p_{3}^2 x_{3} (x_{4}
   x_{5}+x_{4} x_{6}+x_{4} x_{7}+x_{4} x_{8}+x_{7}
   x_{8}) \cr
   +p_{4}^2 x_{6} (x_{1} x_{7}+x_{2} x_{7}+x_{3}
   x_{7}+x_{4} x_{7}+x_{4} x_{8}+x_{7} x_{8})
   +p_{5}^2
   x_{5} x_{6} (x_{1}+x_{2}+x_{3}+x_{4}+x_{8}) \cr
   +p_{6}^2
   x_{5} x_{8} (x_{1}+x_{2}+x_{3}+x_{4})+p_{7}^2 x_{8}
   (x_{2}+x_{3}+x_{4}) (x_{5}+x_{6}+x_{7})\cr
   +(p_1+p_2)^2 x_{3}
   (x_{1} x_{5}+x_{1} x_{6}+x_{1} x_{7}+x_{1}
   x_{8}+x_{5} x_{8}+x_{6} x_{8}+x_{7} x_{8})\cr
   +(p_1+p_2+p_3)^2
   (x_{1} x_{4} x_{5}+x_{1} x_{4} x_{6}+x_{1} x_{4}
   x_{7}+x_{1} x_{4} x_{8}-x_{2} x_{7} x_{8}-x_{3}
   x_{7} x_{8}+x_{4} x_{5} x_{8}+x_{4} x_{6}
   x_{8})\cr
   +(p_2+p_3)^2 x_{2} (x_{4} x_{5}+x_{4} x_{6}+x_{4}
   x_{7}+x_{4} x_{8}+x_{7} x_{8})+(p_2+p_3+p_4)^2 x_{2} x_{6}
   x_{8}+(p_3+p_4)^2 x_{3} x_{6} x_{8}\cr
   +(p_3+p_4+p_5)^2 x_{3} x_{5}
   x_{8}+(p_4+p_5)^2 x_{5} (x_{1} x_{7}+x_{2} x_{7}+x_{3}
   x_{7}+x_{4} x_{7}+x_{4} x_{8}+x_{7} x_{8})\cr
   -(p_4+p_5+p_6)^2
   x_{8} (-x_{1} x_{7}-x_{2} x_{7}-x_{3} x_{7}+x_{4}
   x_{5}+x_{4} x_{6})+(p_5+p_6)^2 x_{6} x_{8}
   (x_{1}+x_{2}+x_{3}+x_{4})\cr
   -(p_5+p_6+p_7)^2 x_{6} x_{8}
   (x_{2}+x_{3}+x_{4})-(p_6+p_7)^2 x_{5} x_{8}
   (x_{2}+x_{3}+x_{4})+(p_6+p_7+p_1)^2 x_{2} x_{5} x_{8}\cr
   -(p_7+p_1)^2
   x_{2} x_{8} (x_{5}+x_{6}+x_{7})-(p_7+p_1+p_2)^2 x_{3} x_{8}
   (x_{5}+x_{6}+x_{7}),
 \end{multline}
 with $\{p_1,\dots,p_7\}\in\mathbb R^{1,3}$ and $p_1+\cdots +p_7=0$.

 For the numerical cases we studied, we  find that the  five-point  massive pentabox integrals in
figure~\ref{fig:pentabox5pt} has a Picard--Fuchs operator of order 2
and degree 18,
whereas the seven-point  massive pentabox integrals in
figure~\ref{fig:pentabox7pt} has a Picard--Fuchs operator of order
4 and degree 67. We have checked with the factorisation algorithm~\cite{facto} that these Picard--Fuchs operators are
irreducible. We see now a transition in the order of the differential
operator when changing the number of external legs.

Near $t=0$, the indicial equations are $\rho(\rho+1)=0$ for the  five-point
case and $(\rho+1)\rho(\rho-1)(\rho-2)=0$,
for the  seven-point case, therefore  a canonical local  basis of solutions behaves for the
five-point case as $t^r$ with $r=-1,0$,
and for the seven-point case as $t^r$  with $r=-1,0,1,2$.
We have checked  in both case the differential operators have only analytic
solution near $t=0$.  In the five-point case for  the numerical
studied, {\tt Maple} identifies the differential equation as being
Louvillian. These results are given on the page~\href{https://nbviewer.org/github/pierrevanhove/PicardFuchs/blob/main/PF-Pentabox.ipynb}{PF-Pentabox}.

This will be proven using a Hodge theoretic analysis for generic
physical parameters in~\cite{DHV}.

%%%%%%%%%%%%%%%%%%%%%%%%%%%%%%%%%%%%%%%%%%%%%%%%%%%%%%%%%%%%%%%%%
\section{Conclusion}\label{sec:conclusion}

In this work, we have used  the algorithm
of~\cite{Lairez} for deriving the Picard--Fuchs operator for rational
differential form in $\mathbb P^{n-1}$
\begin{equation}
\Omega_n(t)=   {\mathcal U(\underline x)^{n-(L+1)D/2}\over (
   \mathcal U(\underline x) \mathcal L_n(\underline m^2,\underline x)-
   t \mathcal V(\underline s,\underline
   x))^{n-LD/2}}\Omega_0^{(n)} ,
\end{equation}
with $D=2$ or $D=4$.
The integration of such  differential form over the positive orthant 
gives the Feynman
integrals arising in many physical problems.
We have presented the differential operator with respect to
the $t$ parameters multiplying the kinematic graph polynomial
$\mathcal V(\underline s,\underline
x)$, but we could have performed the same analysis by
considering the differential operator with respect to any of the
kinematic coefficient entering the coefficient of the monomials of
$\mathcal V(\underline s,\underline
   x)$ or with respect to any of the internal mass $\underline m^2$ in
   the mass hyperplane $\mathcal L(\underline m^2,\underline x)$. 

This algorithm  is an efficient tool for
deriving  the homogeneous part of the differential equation satisfied
by Feynman integrals, because it spares the computation of the certificates (the $Q$ pieces in~\eqref{e:Q}).
It assists the exploration of the changes in the differential
operator for various
configurations of the physical parameters. These changes reflect a modification in the  number of
periods integrals implied by modification of the relations between the
graph polynomial coefficients.

Our main findings are: (1) to have given  some support to the
conjecture identifying the multi-loop sunset integrals as relative
Calabi--Yau period integrals of dimension $n-2$.
(2)  To have given support to the conjecture that the
generic tardigrade
two-loop integral is a
relative period of $K3$ surface of Picard number 11. (3) That the 
double-box differential operator leads to a second order differential with
solutions with logarithmic monodromies, but for the kite and
the pentabox graph the differential operator has only Liouvillian solutions.
(4) Showed that 
splitting an edge of the
sunset integral to make a multi-scoop ice-cream cone Feynman integral changes
drastically the structure of the differential operator. For instance,
at two-loop order, the maximal cut of sunset integral is a  (relative)
period of an open elliptic
curve, but the maximal cut of the two-loop ice-cream cone graph is the one
of a rational surface. (5) Exhibiting how various kinematic
configurations and the effect of the Gram determinant condition
affects the differential operator.

Special values of the kinematics or the
mass parameters, changes the structure of the graph polynomials by
either providing relation between the monomial or having monomial to
vanish. This clearly affects the number of independent period defined
by the rational differential form and consequently the order of the
Picard--Fuchs operator. The presented algorithm detects these
changes. It is tempting to interpret these different values of the
physical parameters in the language of the geometric transition.

%%%%%%%%%%%%%%%%%%%%%%%%%%%%%%%%%%%%%%%%%%%%%%%%%%%%%%%%%%%%%%%%%

\section*{Acknowledgments}
We thank David Broadhurst, Francis Brown, Charles Doran, Andrew Harder, Andrey Novoseltsev for
discussions and comments.
We specially thank Alexandre Goyer and Marc Mezzarobba
for help in factoring differential operators.
We are grateful to IHES for making their
computer resources available.
This work has been supported by the ANR grant ``Amplitude'' ANR-17-
CE31-0001-01,  the ANR grant ``SMAGP'' ANR-20-CE40-0026-01,
the ANR grant ``De Rerum Natura'' ANR-19-CE40-0018,
and by the European Research Council under the European Union's Horizon Europe research and innovation programme, grant agreement 101040794 (10000~DIGITS).


\begin{thebibliography}{Bl1}
\bibitem{Golubeva} V. A. Golubeva, ``Some Problems In The Analytic
  Theory Of Feynman Integrals'' , Russ. Math. Surv. {\bf 31} 139 (1976)

\bibitem{Pham} F.~Pham, ``Introduction \`a l'\'etude topologique des
  singularit\'es de Landau'', Paris : Gauthier-Villars; 1967


\bibitem{Panzer:2015ida}
E.~Panzer,
``Feynman Integrals and Hyperlogarithms,''
% doi:10.18452/17157
Thesis: PhD Humboldt U. (2015)
[arXiv:1506.07243 [math-ph]].
  
\bibitem{Duhr:2019wtr}
C.~Duhr,
``Function Theory for Multiloop Feynman Integrals,''
Ann. Rev. Nucl. Part. Sci. \textbf{69} (2019), 15-39
%doi:10.1146/annurev-nucl-101918-023551

\bibitem{Mizera:2019ose}
S.~Mizera,
``Status of Intersection Theory and Feynman Integrals,''
PoS \textbf{MA2019} (2019), 016
%doi:10.22323/1.383.0016
[arXiv:2002.10476 [hep-th]].

\bibitem{Broadhurst:1995km}
D.~J.~Broadhurst and D.~Kreimer,
``Knots and Numbers in Ph$ i^4$ Theory to 7 Loops and Beyond,''
Int.\ J.\ Mod.\ Phys.\ C {\bf 6} (1995) 519
[hep-ph/9504352].
%%CITATION = HEP-PH/9504352;%%

\bibitem{Broadhurst:1996kc}
D.~J.~Broadhurst and D.~Kreimer,
``Association of Multiple Zeta Values with Positive Knots via Feynman Diagrams Up to 9 Loops,''
Phys.\ Lett.\ B {\bf 393} (1997) 403
[hep-th/9609128].
%%CITATION = HEP-TH/9609128;%%
  
\bibitem{Kontsevich:2001}
M.~Kontsevich and D.~Zagier, ``Periods'', in Engquist, Bj\"orn; Schmid,
Wilfried, Mathematics unlimited -- 2001 and beyond, Berlin, New York:
Springer-Verlag, pp. 771-808.

\bibitem{Bloch:2005bh}
S.~Bloch, H.~Esnault and D.~Kreimer,
``On Motives associated to graph polynomials,''
Commun. Math. Phys. \textbf{267} (2006), 181-225
%doi:10.1007/s00220-006-0040-2
[arXiv:math/0510011 [math.AG]].

\bibitem{BrownCosmic} F.~C.~S.~Brown, ``Feynman Amplitudes and Cosmic
  Galois group'', [arXiv:1512.06409]


\bibitem{Brown:2009ta}
F.~C.~S.~Brown,
``On the Periods of Some Feynman Integrals,''
[arXiv:0910.0114 [math.AG]].

%\cite{Bloch:2014qca}
\bibitem{Bloch:2014qca}
S.~Bloch, M.~Kerr and P.~Vanhove,
``A Feynman Integral via Higher Normal Functions,''
Compos. Math. \textbf{151} (2015) no.12, 2329-2375
doi:10.1112/S0010437X15007472
[arXiv:1406.2664 [hep-th]].
%134 citations counted in INSPIRE as of 08 Apr 2022

%\cite{Bloch:2016izu}
\bibitem{Bloch:2016izu}
S.~Bloch, M.~Kerr and P.~Vanhove,
``Local Mirror Symmetry and the Sunset Feynman Integral,''
Adv. Theor. Math. Phys. \textbf{21} (2017), 1373-1453
%doi:10.4310/ATMP.2017.v21.n6.a1
[arXiv:1601.08181 [hep-th]].
%109 citations counted in INSPIRE as of 30 Mar 2022

\bibitem{Bourjaily:2018ycu}
J.~L.~Bourjaily, Y.~H.~He, A.~J.~Mcleod, M.~Von Hippel and M.~Wilhelm,
``Traintracks Through Calabi--Yau Manifolds: Scattering Amplitudes Beyond Elliptic Polylogarithms,''
Phys. Rev. Lett. \textbf{121} (2018) no.7, 071603
%doi:10.1103/PhysRevLett.121.071603
[arXiv:1805.09326 [hep-th]].

\bibitem{Bourjaily:2019hmc}
J.~L.~Bourjaily, A.~J.~McLeod, C.~Vergu, M.~Volk, M.~Von Hippel and M.~Wilhelm,
``Embedding Feynman Integral (Calabi--Yau) Geometries in Weighted Projective Space,''
JHEP \textbf{01} (2020), 078
%doi:10.1007/JHEP01(2020)078
[arXiv:1910.01534 [hep-th]].

\bibitem{Bourjaily:2018yfy}
J.~L.~Bourjaily, A.~J.~McLeod, M.~von Hippel and M.~Wilhelm,
``Bounded Collection of Feynman Integral Calabi--Yau Geometries,''
Phys. Rev. Lett. \textbf{122} (2019) no.3, 031601
%doi:10.1103/PhysRevLett.122.031601
[arXiv:1810.07689 [hep-th]].

\bibitem{Klemm:2019dbm}
A.~Klemm, C.~Nega and R.~Safari,
``The $l$-loop Banana Amplitude from Gkz Systems and Relative Calabi--Yau Periods,''
JHEP \textbf{04} (2020), 088
%doi:10.1007/JHEP04(2020)088
[arXiv:1912.06201 [hep-th]].

\bibitem{Bonisch:2020qmm}
K.~B\"onisch, F.~Fischbach, A.~Klemm, C.~Nega and R.~Safari,
``Analytic Structure of All Loop Banana Integrals,''
JHEP \textbf{05} (2021), 066
doi:10.1007/JHEP05(2021)066
[arXiv:2008.10574 [hep-th]].

\bibitem{Bonisch:2021yfw}
K.~B\"onisch, C.~Duhr, F.~Fischbach, A.~Klemm and C.~Nega,
``Feynman Integrals in Dimensional Regularization and Extensions of Calabi--Yau Motives,''
[arXiv:2108.05310 [hep-th]].

\bibitem{Bourjaily:2022bwx}
J.~L.~Bourjaily, J.~Broedel, E.~Chaubey, C.~Duhr, H.~Frellesvig, M.~Hidding, R.~Marzucca, A.~J.~McLeod, M.~Spradlin and L.~Tancredi, \textit{et al.}
``Functions Beyond Multiple Polylogarithms for Precision Collider Physics,''
[arXiv:2203.07088 [hep-ph]].
  
  \bibitem{Forum:2022lpz}
A.~Forum and M.~von Hippel,
``A Symbol and Coaction for Higher-Loop Sunrise Integrals,''
[arXiv:2209.03922 [hep-th]].

\bibitem{Duhr:2022pch}
C.~Duhr, A.~Klemm, F.~Loebbert, C.~Nega and F.~Porkert,
``Yangian-Invariant Fishnet Integrals in 2 Dimensions as Volumes of Calabi--Yau Varieties,''
[arXiv:2209.05291 [hep-th]].

\bibitem{Vanhove:2014wqa}
P.~Vanhove,
``The Physics and the Mixed Hodge Structure of Feynman Integrals,''
Proc. Symp. Pure Math. \textbf{88} (2014), 161-194
%doi:10.1090/pspum/088/01455
[arXiv:1401.6438 [hep-th]].

\bibitem{facto}  F.~Chyzak, A.~Goyer, and M.~Mezzarobba,
  ``Symbolic-Numeric Factorization of Differential Operators'', [arXiv:2205.08991]


\bibitem{VanhoveISSAC} P. Vanhove  ``Differential Equations for Feynman Integrals.'' Proceedings of the 2021 on International Symposium on Symbolic and Algebraic Computation, 21-26. https://doi.org/10.1145/3452143.3465512
  
\bibitem{Vanhove:2018mto}
P.~Vanhove,
``Feynman Integrals, Toric Geometry and Mirror Symmetry,''
%doi:10.1007/978-3-030-04480-0\_17
[arXiv:1807.11466 [hep-th]].

 \bibitem{Lairez} P.~Lairez, ``Computing periods of rational
   integrals'', Math. Comp. {\bf 85} (2016), 1719-1752, [arXiv:1404.5069]


\bibitem{Bitoun:2017nre}
T.~Bitoun, C.~Bogner, R.~P.~Klausen and E.~Panzer,
``Feynman Integral Relations from Parametric Annihilators,''
Lett. Math. Phys. \textbf{109} (2019) no.3, 497-564
%doi:10.1007/s11005-018-1114-8
[arXiv:1712.09215 [hep-th]].

  \bibitem{Nakanishi}  Noboru Nakanishi, ``Graph Theory and Feynman Integrals'', Gordon \& Breach Science Publishers Ltd (1971)

\bibitem{Itzykson:1980rh}
C.~Itzykson and J.~B.~Zuber,``Quantum Field Theory,'' McGraw-Hill,  New York, 1980

  
\bibitem{Bogner:2010kv}
C.~Bogner and S.~Weinzierl,
``Feynman Graph Polynomials,''
Int. J. Mod. Phys. A \textbf{25} (2010), 2585-2618
%doi:10.1142/S0217751X10049438
[arXiv:1002.3458 [hep-ph]].

\bibitem{Weinzierl:2022eaz}
S.~Weinzierl,
``Feynman Integrals,''
[arXiv:2201.03593 [hep-th]].

%\cite{Asribekov:1962tgp}
\bibitem{Asribekov:1962tgp}
V.~E.~Asribekov,
``Choice of Invariant Variables for the `'Many-Point'' Functions,''
J. Exp. Theor. Phys. \textbf{15} (1962) no.2, 394
%4 citations counted in INSPIRE as of 30 Mar 2022
  
\bibitem{Eden:1966dnq}
R.~J.~Eden, P.~V.~Landshoff, D.~I.~Olive and J.~C.~Polkinghorne,
``The analytic S-matrix,''
Cambridge University Press, 2002.

 \bibitem{Hannesdottir:2022bmo}
H.~S.~Hannesdottir and S.~Mizera,
``What is the $i\varepsilon$ for the S-Matrix?,''
[arXiv:2204.02988 [hep-th]]. 


\bibitem{Weinberg:1959nj}
S.~Weinberg,
``High-Energy Behavior in Quantum Field Theory,''
Phys. Rev. \textbf{118} (1960), 838-849
%doi:10.1103/PhysRev.118.838


\bibitem{Speer:1975dc}
E.~R.~Speer,
``Ultraviolet and Infrared Singularity Structure of Generic Feynman Amplitudes,''
Ann. Inst. H. Poincare Phys. Theor. \textbf{23} (1975), 1-21


\bibitem{Speer}  E. R. Speer, ``Generalized Feynman Amplitudes,'' vol. 62 of Annals of Mathematics Studies. Princeton University Press, New Jersey, Apr., 1969.

  
\bibitem{Laporta:2000dc}
S.~Laporta,
``Calculation of Master Integrals by Difference Equations,''
Phys. Lett. B \textbf{504} (2001), 188-194
%doi:10.1016/S0370-2693(01)00256-8
[arXiv:hep-ph/0102032 [hep-ph]].

\bibitem{Smirnov:2010hn}
A.~V.~Smirnov and A.~V.~Petukhov,
``The Number of Master Integrals is Finite,''
Lett. Math. Phys. \textbf{97} (2011), 37-44
%doi:10.1007/s11005-010-0450-0
[arXiv:1004.4199 [hep-th]].

\bibitem{Lee:2013hzt}
R.~N.~Lee and A.~A.~Pomeransky,
``Critical Points and Number of Master Integrals,''
JHEP \textbf{11} (2013), 165
%doi:10.1007/JHEP11(2013)165
[arXiv:1308.6676 [hep-ph]].

\bibitem{Henn:2014qga}
J.~M.~Henn,
``Lectures on differential equations for Feynman integrals,''
J. Phys. A \textbf{48} (2015), 153001
%doi:10.1088/1751-8113/48/15/153001
[arXiv:1412.2296 [hep-ph]].
  
\bibitem{GKZ} I. M. Gelfand, M. M. Kapranov,~ And A. V. Zelevinsky,
  ``Generalized Euler Integrals and A-Hypergeometric Functions'',
  Advances In Mathematics {\bf 84}, 255-271 (1990).
  
\bibitem{Klausen:2019hrg}
R.~P.~Klausen,
``Hypergeometric Series Representations of Feynman Integrals by Gkz Hypergeometric Systems,''
JHEP \textbf{04} (2020), 121
%doi:10.1007/JHEP04(2020)121
[arXiv:1910.08651 [hep-th]].

\bibitem{Feng:2019bdx}
T.~F.~Feng, C.~H.~Chang, J.~B.~Chen and H.~B.~Zhang,
``Gkz-Hypergeometric Systems for Feynman Integrals,''
Nucl. Phys. B \textbf{953} (2020), 114952
%doi:10.1016/j.nuclphysb.2020.114952
[arXiv:1912.01726 [hep-th]].


\bibitem{delaCruz:2019skx}
  L.~de la Cruz,
 ``Feynman Integrals as A-Hypergeometric Functions,''
JHEP \textbf{12} (2019), 123
%doi:10.1007/JHEP12(2019)123
[arXiv:1907.00507 [math-ph]].

\bibitem{Tarasov:1996br}
  O.~V.~Tarasov,
  ``Connection Between Feynman Integrals Having Different Values of the Space-Time Dimension,''
  Phys.\ Rev.\ D {\bf 54} (1996) 6479
  [hep-th/9606018].

  
\bibitem{Koutchan} C. Koutschan. ``HolonomicFunctions (user’s guide).'' Technical Report 10-01, RISC Report Series, Johannes Kepler University, Linz, Austria, 2010. http://www.risc.jku.at/research/ combinat/software/HolonomicFunctions/.
  
\bibitem{bostan2013creative} A. Bostan, P. Lairez, and B. Salvy, ''Creative telescoping for rational functions using the Griffiths--Dwork method.'' In Proceedings of the 38th international symposium on symbolic and algebraic computation (pp. 93-100).
  
\bibitem{Picard1899} \'E. Picard. ``Quelques remarques sur les int\'egrales doubles de seconde esp\`ece dans la th\'eorie des surfaces alg\'ebriques.'', \emph{C. R. Acad. Sci. Paris}, 129:539–540, 1899.

\bibitem{Griffiths_1969}
P.~A. Griffiths.
\newblock On the periods of certain rational integrals.
\newblock {Ann. of Math.}, {\bf 90} (1969), 460--541.

\bibitem{Dwork_1962}
B.~Dwork.
\newblock On the zeta function of a hypersurface.
\newblock {Inst. Hautes \'Etudes Sci. Publ. Math.} {\bf 12} (1962) 5--68.

\bibitem{Dwork_1964}
B.~Dwork.
\newblock On the zeta function of a hypersurface: {{II}}.
\newblock {Ann. of Math.}, {\bf 80} (1964) 227--299.

\bibitem{Verrill}H. Verrill, \emph{Root lattices and pencils
of varieties}, J. Math. Kyoto Univ. 36 (2) (1996), 423-446.

  
\bibitem{Batyrev:1998kx}
V.~V.~Batyrev, I.~Ciocan-Fontanine, B.~Kim and D.~van Straten,
``Conifold transitions and mirror symmetry for Calabi--Yau complete intersections in Grassmannians,''
Nucl. Phys. B \textbf{514} (1998), 640-666
%doi:10.1016/S0550-3213(98)00020-0
[arXiv:alg-geom/9710022 [math.AG]].

\bibitem{Hori:2000kt}
K.~Hori and C.~Vafa,
``Mirror symmetry,''
[arXiv:hep-th/0002222 [hep-th]].
  
\bibitem{Coates} T.~Coates, A.~Corti, S.~Galkin,  V.~Golyshev,  and
  A.~Kasprzyk, (2012). ``Mirror symmetry and Fano manifolds.''
  European Congress of Mathematics (Krak{\'o}w, 2-7 July, 2012),
  November 2013, pp.\ 285--300 [arXiv:1212.1722.]


  %\cite{Bloch:2013tra}
\bibitem{Bloch:2013tra}
S.~Bloch and P.~Vanhove,
``The Elliptic Dilogarithm for the Sunset Graph,''
J. Number Theor. \textbf{148} (2015), 328-364
%doi:10.1016/j.jnt.2014.09.032
[arXiv:1309.5865 [hep-th]].
%190 citations counted in INSPIRE as of 11 Apr 2022

  
\bibitem{DNV} C.~Doran,  A.~Novoseltsev and P.~Vanhove, ``Mirroring
  Towers: \\The Calabi--Yau Geometry of the Multiloop Sunset Feynman
  Integrals'' to appear.
  
\bibitem{Candelas:2021lkc}
P.~Candelas, X.~de la Ossa, P.~Kuusela and J.~McGovern,
``Mirror Symmetry for Five-Parameter Hulek-Verrill Manifolds,''
[arXiv:2111.02440 [hep-th]].

\bibitem{MullerStach:2012mp}
  S.~M\"uller-Stach, S.~Weinzierl and R.~Zayadeh,
  ``Picard-Fuchs Equations for Feynman Integrals,''
  Commun.\ Math.\ Phys.\  {\bf 326} (2014) 237
 % doi:10.1007/s00220-013-1838-3
  [arXiv:1212.4389 [hep-ph]].
  
\bibitem{Kreimer:2022fxm}
D.~Kreimer,
``Bananas: multi-edge graphs and their Feynman integrals,''
[arXiv:2202.05490 [hep-th]].


\bibitem{MullerStach:2011qkg}
S.~M\"uller-Stach, S.~Weinzierl and R.~Zayadeh,
``A Second-Order Differential Equation for the Two-Loop Sunrise Graph with Arbitrary Masses,''
Commun. Num. Theor. Phys. \textbf{6} (2012), 203-222
%doi:10.4310/CNTP.2012.v6.n1.a5
[arXiv:1112.4360 [hep-ph]].

 \bibitem{VanhoveStringMath2019} P.~Vanhove,''Mirroring towers of Feynman integrals: Fibration and degeneration in Feynman integral Calabi--Yau geometries'', (String Math 2019)

\bibitem{Verrill3} H. Verrill, \emph{Sums of squares of binomial
    coefficients, with applications to Picard--Fuchs equations}, [arXiv:math/0407327]
  
\bibitem{ore} M. Kauers, M. Jaroschek, F.  Johansson, ``Ore
  Polynomials In Sage'', {\tt
    http://www.risc.jku.at/research/combinat/software/ore\_algebra}, [arXiv 1306.4263]

\bibitem{ore2} M. Mezzarobba, ``Rigorous Multiple-Precision Evaluation
  of D-Finite Functions in SageMath'',5th International Congress on
  Mathematical Software (ICMS~2016), Jul 2016, Berlin, Germany,  [arXiv:1607.01967]
  
\bibitem{Klausen:2021yrt}
R.~P.~Klausen,
``Kinematic singularities of Feynman integrals and principal A-determinants,''
JHEP \textbf{02} (2022), 004
%doi:10.1007/JHEP02(2022)004
[arXiv:2109.07584 [hep-th]].


  
\bibitem{Fakler}  W.~Fakler, ``On second order homogeneous linear
  differential equations with Liouvillian solutions'', Theoretical Computer Science 187 (1997) 27-48

  
\bibitem{DHV}
C.~F.~Doran, A.~Harder, E.~Pichon-Pharabod and P.~Vanhove,
``Motivic Geometry of Two-Loop Feynman Integrals,''
[arXiv:2302.14840 [math.AG]].


  
\bibitem{Morrison:1991cd}
  D.~R.~Morrison,
  ``Picard-Fuchs Equations and Mirror Maps for Hypersurfaces,''
  AMS/IP Stud.\ Adv.\ Math.\  {\bf 9} (1998) 185
  [hep-th/9111025].

\bibitem{Duhr:2022dxb}
C.~Duhr, A.~Klemm, C.~Nega and L.~Tancredi,
``The Ice Cone Family and Iterated Integrals for Calabi-Yau Varieties,''
JHEP \textbf{02} (2023), 228
[arXiv:2212.09550 [hep-th]].

  
\bibitem{Broadhurst:1987ei}
D.~J.~Broadhurst,
``The Master Two Loop Diagram With Masses,''
Z. Phys. C \textbf{47} (1990), 115-124
%doi:10.1007/BF01551921
  
\bibitem{Remiddi:2016gno}
E.~Remiddi and L.~Tancredi,
``Differential Equations and Dispersion Relations for Feynman Amplitudes. the Two-Loop Massive Sunrise and the kite Integral,''
Nucl. Phys. B \textbf{907} (2016), 400-444
%doi:10.1016/j.nuclphysb.2016.04.013
[arXiv:1602.01481 [hep-ph]].
  
\bibitem{Adams:2016xah}
L.~Adams, C.~Bogner, A.~Schweitzer and S.~Weinzierl,
``The kite Integral to All Orders in Terms of Elliptic Polylogarithms,''
J. Math. Phys. \textbf{57} (2016) no.12, 122302
%doi:10.1063/1.4969060
[arXiv:1607.01571 [hep-ph]].

\bibitem{Bogner:2017vim}
C.~Bogner, A.~Schweitzer and S.~Weinzierl,
``Analytic Continuation and Numerical Evaluation of the kite Integral and the Equal Mass Sunrise Integral,''
Nucl. Phys. B \textbf{922} (2017), 528-550
%doi:10.1016/j.nuclphysb.2017.07.008
[arXiv:1705.08952 [hep-ph]].

\bibitem{Bogner:2018uus}
C.~Bogner, A.~Schweitzer and S.~Weinzierl,
``Analytic Continuation of the kite Family,''
%doi:10.1007/978-3-030-04480-0\_4
[arXiv:1807.02542 [hep-th]].


\bibitem{Broedel:2018qkq}
J.~Broedel, C.~Duhr, F.~Dulat, B.~Penante and L.~Tancredi,
``Elliptic Feynman Integrals and Pure Functions,''
JHEP \textbf{01} (2019), 023
doi:10.1007/JHEP01(2019)023
[arXiv:1809.10698 [hep-th]].


\bibitem{Bezuglov:2020ywm}
M.~A.~Bezuglov, A.~I.~Onishchenko and O.~L.~Veretin,
``Massive kite Diagrams with Elliptics,''
Nucl. Phys. B \textbf{963} (2021), 115302
%doi:10.1016/j.nuclphysb.2020.115302
[arXiv:2011.13337 [hep-ph]].
  
\bibitem{Caron-Huot:2012awx}
S.~Caron-Huot and K.~J.~Larsen,
``Uniqueness of Two-Loop Master Contours,''
JHEP \textbf{10} (2012), 026
%doi:10.1007/JHEP10(2012)026
[arXiv:1205.0801 [hep-ph]].

%\cite{Bloch:2021hzs}
\bibitem{Bloch:2021hzs}
S.~Bloch,
``Double Box Motive,''
SIGMA \textbf{17} (2021), 048
%doi:10.3842/SIGMA.2021.048
[arXiv:2105.06132 [math.AG]].
%3 citations counted in INSPIRE as of 30 Mar 2022

%\cite{Bourjaily:2017bsb}
\bibitem{Bourjaily:2017bsb}
J.~L.~Bourjaily, A.~J.~McLeod, M.~Spradlin, M.~von Hippel and M.~Wilhelm,
``Elliptic Double-Box Integrals: Massless Scattering Amplitudes Beyond Polylogarithms,''
Phys. Rev. Lett. \textbf{120} (2018) no.12, 121603
%doi:10.1103/PhysRevLett.120.121603
[arXiv:1712.02785 [hep-th]].
%111 citations counted in INSPIRE as of 30 Mar 2022

\bibitem{Pozo:2022dox}
A.~C.~Pozo and M.~von Hippel,
``A Three-Parameter Elliptic Double-Box,''
[arXiv:2209.03921 [hep-th]].
  
\end{thebibliography}
\end{document}